\documentclass[prd,aps,a4paper,eqsecnum,floatfix,nofootinbib,twocolumn]{revtex4-2}  

\usepackage{graphicx} 
\usepackage{amsmath,amsfonts,amssymb}
\usepackage{placeins}
\usepackage{tikz}
\usepackage{appendix}
\usepackage[caption=false,justification=justified]{subfig}

\newcommand{\beq}{\begin{equation}}
\newcommand{\eeq}{\end{equation}}
\newcommand{\bea}{\begin{eqnarray}}
\newcommand{\eea}{\end{eqnarray}}
\newcommand{\be}{\begin{equation}}      
\newcommand{\ee}{\end{equation}}
\def\nn{\nonumber}

\begin{document}
\title{Scalar radiation from circular orbits in D3 brane spacetimes: a  Mano-Suzuki-Takasugi-like approach}

\author{Donato Bini$^{1}$, Giorgio Di Russo$^{2}$, Veronica Fantini$^3$}   
  \affiliation{
$^1$Istituto per le Applicazioni del Calcolo ``M. Picone,'' CNR, I-00185 Rome, Italy\\
$^2$School of Fundamental Physics and Mathematical Sciences, Hangzhou Institute for Advanced Study, UCAS, Hangzhou 310024, China\\
$^3$Universit\'e Paris-Saclay, CNRS, Laboratoire de Mathematique d'Orsay (LMO), 91405 Orsay, France
}

\date{\today}

\begin{abstract}
We study scalar  perturbations in a ten-dimensional D3 brane spacetime, developing a (higher-dimensional) Mano-Suzuki-Takasugi-like formalism adapted to the present case leading to a sourced modification of a Mathieu equation. As an application, we compute the energy loss both at infinity and at the horizon in terms of scalar waves due to a scalar charge   moving along a circular \lq\lq equatorial" geodesic.
\end{abstract}

\maketitle 

\section{Introduction}
D-branes play a central role in string theory, providing a natural link between the perturbative description of strings and the low-energy dynamics of gravity and gauge fields \cite{Polchinski:1996na}. Among the various types of D-branes in type II string theory, D3 branes are of particular interest. They are intrinsically related to type IIB superstring theory, preserve one half of the supersymmetries, and source a self-dual Ramond--Ramond five-form field strength\cite{Polchinski:1995mt}. Moreover, their near-horizon geometry provides one of the most important realizations of the gauge/gravity correspondence, through the emergence of an $AdS_5\times S^5$ spacetime\cite{Maldacena:1997re}.

A stack of  $N$ coincident D3 branes is described, in the low-energy regime, by a ten-dimensional solution of type IIB supergravity. In the string frame, the corresponding geometry can be written as
\bea
\label{metr}
ds^2
&=&
H(r)^{-1/2}
\left(
-c^2 dt^2+d\mathbf{x}^2
\right)\nonumber\\
&+&
H(r)^{1/2}
\left(
dr^2+r^2d\Omega_5^2
\right),
\eea
where $\mathbf{x}=(x^1,x^2,x^3)$ denotes the spatial directions parallel to the branes and
\beq
\label{accar}
H(r)=1+\frac{L^4}{r^4}\,,
\eeq
and $L$ a parameter with the dimensions of a length, representing the charge of the D3 brane.
$H(r)$ is a harmonic function in $d=6$, in the sense that it satisfies the (radial) Laplace equation
\beq
\left(\partial_{rr}+\frac{d-1}{r}\partial_r \right) H(r)=0\,,
\eeq
for $d=6$.
Moreover, the angular element (depending on the angles $\theta_i=\{\theta_1, \theta_2, \theta_3, \theta_4, \theta_5\}$) reads
\beq
d\Omega_5^2=d\theta^2+\sin^2\theta d\Omega_4^2\,,
\eeq
or explicitly
\bea
d\Omega_5^2 &=&
d\theta_1^2
+\sin^2\theta_1
\left[
d\theta_2^2
+\sin^2\theta_2
\left(
d\theta_3^2 \right.\right.\nonumber\\
&+&\left.\left.\sin^2\theta_3
\left(
d\theta_4^2
+\sin^2\theta_4\,d\theta_5^2
\right)
\right)
\right]\,,
\eea
where $\theta_1=\theta$ and $\theta_5=\phi$ is  also a used notation to make contact with the standard four dimensional spacetime of classical General Relativity (GR).  Hereafter we will work  in units of $c=1$.
The metric \eqref{metr} admits a horizon located at $r=0$ and is characterized by a vanishing 
Ricci scalar. In addition
\beq
\sqrt{-g}=\sqrt{H(r)} r^5\sin^4(\theta_1)\sin^3(\theta_2)\sin^2(\theta_3)\sin(\theta_4)\,. 
\eeq

The characteristic length scale $L$ is determined by the string parameters and the string coupling. Indeed, in conventional string units one has
\beq
L^4=4\pi g_s N\,\alpha'^2,
\eeq
with $N$ the number of branes in the stack, $g_s$ the string coupling constant such that the ten-dimensional Newton constant is related to the string parameters according to
\beq
G_{10}
=
\frac{(2\pi)^7}{16\pi}\,
g_s^2\,\alpha'^4 c^8
=
8\pi^6 g_s^2\alpha'^4 c^8\,.
\eeq
Equivalently, the scale $L$ can be expressed directly in terms of $G_{10}$, $N$, and $g_s$. These relations make explicit the connection between the microscopic string description and the macroscopic geometry generated by the branes.

The remarkable properties of the D3 brane solution become particularly transparent in the near-horizon region
\beq
r\ll L\,.
\eeq
In this limit  $H(r)\simeq L^4/r^4$, and the metric approaches 
\beq
ds^2
\simeq
\frac{r^2}{L^2}
\left(
-dt^2+d\mathbf{x}^2
\right)
+
\frac{L^2}{r^2}dr^2
+
L^2d\Omega_5^2.
\eeq
Thus, up to the appropriate choice of coordinates, the geometry becomes
$AdS_5\times S^5$.
Such a geometry underlies the celebrated correspondence between type IIB string theory on \(AdS_5\times S^5\) and four-dimensional gauge dynamics on the world-volume of the D3 branes.

The perturbative dynamics of D-branes and their associated supergravity backgrounds have been extensively investigated, with particular attention to the stability of these configurations and their response to external perturbations. In this context, quasinormal modes (QNMs), which characterize the damped oscillations of perturbed gravitational systems, provide a valuable tool for probing their dynamical properties and stability \cite{Berti:2009kk,Bianchi:2021xpr,Bianchi:2022qph}.

D3 brane spacetimes have been also studied as examples of spacetimes undergoing the Couch–Torrence (CT) symmetry.
Indeed, a remarkable symmetry of extremal BH geometries is the Couch–Torrence (CT) conformal inversion, originally discovered for the extremal Reissner–Nordstr\"om (RN) solution \cite{CouchTorrence1984}. This discrete transformation combines a radial inversion with a conformal rescaling and effectively exchanges the near-horizon region with the asymptotic region, revealing a nontrivial connection between the dynamics of perturbations in these two regimes. Beyond the original RN setting, generalized CT inversions have been identified in higher-dimensional gravitational backgrounds and extended to D-brane geometries \cite{Bianchi:2021yqs}, including rotating D3 branes \cite{Bianchi:2022wku} for suitable choices of their angular momenta. CT conformal inversion symmetries have also been generalized to four-dimensional extremal black holes in $\mathcal{N}=2$ STU supergravity, including the BPS solutions with general asymptotic scalar fields, for which the inversion maps a given BH solution to another member of the same family with transformed parameters \cite{Cvetic:2020kwf,Cvetic:2021lss}. These extensions suggest that conformal inversion symmetries provide a useful framework for uncovering hidden relations between the near-horizon and asymptotic dynamics of BHs and branes. The latter are also addressed here for the specific case under consideration, see below.

The main goal of this paper is to compute the scalar wave  flux generated by a massive point particle moving along circular orbits in a D-brane geometry, within the gravitational self-force framework and using a post-Newtonian (PN) expansion \cite{Blanchet:2013haa}. This analysis will provide a first step towards the investigation of more general orbital configurations and dynamical processes in these backgrounds where the weak field and slow motion  approximation can be relaxed. A crucial ingredient in this program is the construction of solutions to the radial perturbation equation satisfying ingoing boundary conditions at the event horizon and outgoing boundary conditions at infinity. For the D3 brane background of interest here, the radial equation can be reduced to Mathieu form, whose solutions are closely related to particular cases of the confluent Heun equation \cite{Consoli:2022eey}. Exploiting the mathematical properties of Mathieu functions and their connection to the broader Heun family may therefore provide an effective framework for constructing the required boundary-condition-adapted solutions and extracting the PN expansion of the energy flux.

To construct solutions satisfying the required boundary conditions, we will adapt the celebrated Mano–Suzuki–Takasugi (MST) method \cite{Mano:1996vt,Sasaki:2003xr}, originally developed to obtain analytic solutions of the radial Teukolsky equation in black hole spacetimes. Recently, this formalism has been adapted to scalar perturbations of five-dimensional topological star (TS) and W soliton geometries\cite{Bianchi:2024vmi,Bianchi:2024rod,DiRusso:2025lip,Bianchi:2025aei,Bianchi:2025ydq} and extended to massless scalar perturbations of the five-dimensional Schwarzschild–Tangherlini \cite{Bini:2026foz} spacetime through an MST-like approach based on the Reduced Confluent Heun equation.

Recently, alternative methods to MST have been developed to solve Heun equations within specific approximation schemes. One is based on instanton calculus and the correspondence between Heun equations and Seiberg–Witten curves \cite{Cipriani:2025ikx,Cipriani:2026myb}, while the other employs the kink method, which connects Floquet solutions around different singular points \cite{Fioravanti:2025bts}.

The plan of the present paper is the following.

Sec. I-III, after a general introduction to D3 brane spacetimes, contain a brief reminder on the associated timelike geodesics (including circular orbits, scattering orbits  and radial infalling orbits). In Sec. IV  we discuss the Couch-Torrence symmetry and in Sec. V we consider the source-free scalar waves, exploring for them the radial equation with the standard two-types of solutions: PN type (Sec. VI) and MST-type (Sec. VII, including the computation of the renormalized angular momentum parameter as well as its eikonal-type resummation at the NLO). In Sec. VIII we will study the scalar waves sourced by a massive particle moving along a circular timelike geodesic of the background, computing   the radiated energy at infinity, with consequent extraction of gauge-invariant information. 

Notably, in Sec. VII.D, we start a mathematical treatment of the solutions to the radial equation close to the singular points, when all tools associated with weak field analysis are no longer valid. This includes the study of the
divergent formal solutions at the singular points using
the techniques of Borel summability. 

Finally, in Sec. IX, we discuss   possible directions for future work.

\section{Geometrical characterization of the D3 brane spacetimes}

When restricted to the hypersurface $x^i=x_0^i$ (constant) and $\theta_1=\theta_2=\theta_3=\theta_4=\pi/2$  (constant), as will often be the case below, 
the above metric reduces to its \lq\lq equatorial" form
\bea
\label{eq_met}
ds_{\rm eq}^2&=&g^{\rm eq}_{AB}dx^A dx^B\nonumber\\
&=&
-\frac{dt^2}{H(r)^{1/2}}+H(r)^{1/2}\left(dr^2+r^2d\phi^2\right)\,,\qquad
\eea
with $\phi=\theta_5$.
The latter sectional metric is not Ricci flat, i.e., 
\bea
R_{AB}&=&{\rm diag}\left[\frac{r^2 L^4 (L^4-4r^4)}{(r^4+L^4)^3},-\frac{L^4 (L^4-r^4)}{r^2 (r^4+L^4)^2},\right.\nonumber\\
&-&\left. 5\frac{r^4 L^4}{ (r^4+L^4)^2}  \right]\,,
\eea
and the scalar curvature tensor is given by
\beq
R_{\rm eq}=-\frac{2 L^8}{(r^4+L^4)^{5/2}}\,.
\eeq
For $r\to \infty$ then 
\beq
L^2 R_{\rm eq}\approx -2\left(\frac{L}{r}\right)^{10} +O\left(\frac{1}{r^{14}}\right)\,,
\eeq
whereas when $r\to L$
\beq
L^2 R_{\rm eq}=-\frac{1}{2\sqrt{2}} \,.
\eeq
Finally, when $r\to 0$
\beq
L^2 R_{\rm eq}=-2 \,.
\eeq
The sign-reversed rescaled Ricci scalar (also depending on conventions) 
\beq
R_{\rm resc}=\frac{2}{\left(1+\frac{r^4}{L^4} \right)^{5/2}}\,, 
\eeq
interpolates then between the values $R_{\rm resc}=2$ (at $r=0$) and $R_{\rm resc}=0$ (at $r\to\infty$), approaching these values with a horizontal tangent. An inflection point exists at $r=L (3/11)^{1/4}$.

\begin{figure}
\includegraphics[scale=0.9]{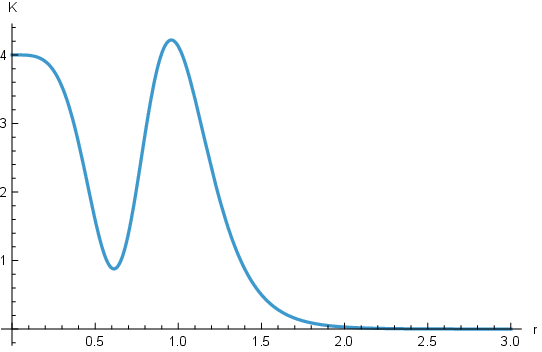}
\caption{\label{fig:1} The Kretschmann scalar $K$ for the sectional 3D metric is shown as a function of $r$ (in units of $L=1$). At the horizon one finds
$K_h=\frac{4}{L^4}$  while  at the light ring, $r=L$,  $K_{\rm (lr)}=\frac{33}{8L^4}\approx \frac{4.125}{L^4}$.  Moreover, the horizon itself corresponds to a maximum and two other extremal points exist at $r\approx 0.6130 L$  (minimum) and $r\approx 0.9582 L$ (maximum). At infinity $K\to 0$.
}
\end{figure}

\begin{figure}
\includegraphics[scale=0.9]{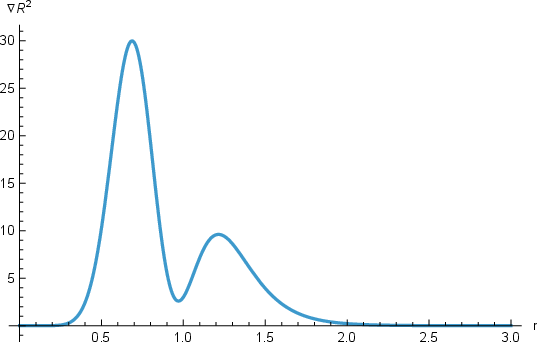}
\caption{\label{fig:2} The scalar $\nabla R^2=R_{\alpha\beta\gamma; \delta}R^{\alpha\beta\gamma; \delta}$  for the sectional 3D metric  is shown as a function of $r$ (in units of $L=1$). At the horizon one finds
$\nabla R^2_h=0$  while  at the light ring $r=L$  $\nabla R^2_{\rm (lr)}=\frac{17\sqrt{2}}{8 L^6}\approx \frac{3.0052}{L^6}$.  
At infinity $\nabla R^2\to 0$. The scalar $\nabla R^2$ is visually nonzero in a very compact region, namely it is practically zero for $r\approx 3 L$ and beyond.
}
\end{figure}

Similarly, the Kretschmann scalar of the metric \eqref{eq_met} is given by
\beq
K=R^{\alpha\beta\gamma\delta}R_{\alpha\beta\gamma\delta}=\frac{4 L^8 (L^8-10 r^4 L^4+42 r^8)}{(r^4+L^4)^5}\,.
\eeq
Its behavior as a function of $r$ (in units of $L=1$) is shown in Fig. \ref{fig:1}.
Fig. \ref{fig:2} shows  instead its covariant derivative analogue, i.e., 
\bea
\nabla R^2&=&R_{\alpha\beta\gamma; \delta}R^{\alpha\beta\gamma; \delta}\nonumber\\
&=&\frac{32 L^8 (198 r^8-330 r^4 L^4+149 L^8) r^8}{ (r^4+L^4)^{15/2}}\,.
\eea

Other useful geometrical and physical information can be obtained by considering the contracted derivative of the Riemann tensor, namely a sort of gravitational current in a gravitoelectromagnetic analogy, say \lq\lq ${\rm div } R=S$", 
\beq
R^{\alpha\beta\gamma\delta}{}_{;\delta}=S^{\alpha\beta\gamma}\,.
\eeq
The latter, after dualization of its first two  (antisymmetric) indices, becomes a 2-tensor
\beq
T^{\alpha\beta}=\eta^\alpha{}_{\mu\nu}S^{\mu\nu \beta}\,,
\eeq
and it is characterized by either a symmetric part $T_S\equiv T^{(\alpha\beta)}$
\beq
T_S=\frac{48 L^4 r^3 (L^4-r^4)}{(r^4+L^4)^{15/4}} [\partial_t \otimes \partial_\phi+\partial_\phi \otimes \partial_t]\,,
\eeq
and an antisymmetric part
$T^{[\alpha\beta]} $ further dualized to give the vector
\beq
T_V\equiv \eta^{\alpha\beta\gamma}T_{\beta\gamma}\partial_\alpha =\frac{20 r^5 L^8}{ (r^4+L^4)^4}\, \partial_r\,,
\eeq
aligned with the radial direction.
Both at the horizon $r=0$ and at the light ring $r=L$ the symmetric 2-tensor $T_S$ vanishes identically. The vector $T_V$ has instead a vanishing magnitude only at the horizon. 
It is worth to repeat that one should consider anyway these as qualitative-type pieces of information, in the sense that the presence of a radial gravitational  current (fast decaying at infinity) is only qualitatively a gravitational source of this spacetime (a section of a D3 brane spacetime) but it might help intuition in certain circumstances, especially to guess the peculiarities of the associated gravitational force.

As already stated, in this paper we study the massless scalar wave equation in the D3 brane spacetime  \eqref{metr}. Over the past decades, string theory has developed a general framework for dealing with problems of this kind, namely the so-called (quantum) Seiberg–Witten formalism \cite{Lerche:1996xu,Aminov:2020yma}. Scalar wave equations arising in the context of black-hole perturbation theory in classical general relativity (GR), on the other hand, have traditionally been studied using the so-called Mano–Suzuki–Takasugi (MST) formalism. Only recently a precise equivalence between these two formalisms has been established.

Moreover, since classical GR is formulated in four spacetime dimensions, extending the MST formalism to dimensions $d>4$ calls for further investigation, in particular to clarify the role played by the dimensionality of the background spacetime. This issue has been explored in specific examples, such as the cases of TS and W-solitons \cite{Bianchi:2024vmi,Bianchi:2024rod,DiRusso:2025lip,Bianchi:2025aei,Bini:2025ltr,Bianchi:2025ydq}, but a more general understanding is still lacking.

In addition, in most of the explicitly analytically considered cases (either in GR and in string theory), one is finally left with the solution of a Heun-type equation (confluent, in the case of black holes, for example). There exist interesting situations where additional confluences arise (reduced confluent or even doubly reduced confluent Heun equation) and it is important to test the formalisms in these contexts too.

Rather than presenting yet another list of closely related applications, we aim to offer both the GR and the self-force communities a set of studies focused on limiting or critical situations, in the hope of contributing to a unified picture that emerges from the various examples.

\section{Geodesic orbits: a brief reminder}

Let us consider massive geodesics in the D3 brane background. For definiteness, we restrict the motion to a two-dimensional plane in the transverse space, so that the metric \eqref{metr} can be written as
\bea
ds^2&=&
H(r)^{-1/2}\left(-dt^2+d\mathbf{x}^2\right)\nonumber\\
&+& H(r)^{1/2}\left(dr^2+r^2d\phi^2\right),
\eea
with $H(r)$ defined in Eq. \eqref{accar} and $\phi\equiv \theta_5$, as already noted.
The geodesic equations admit several first integrals associated with the translational and rotational symmetries of the background. In particular, for a particle of rest mass $\mu$, we introduce the conserved energy $E$, the momentum $\mathbf{p}$ parallel to the brane, and the angular momentum $J$ in the transverse space,
\bea\label{canmom}
E&=&H(r)^{-1/2}\frac{dt}{d\lambda}\,,\nonumber\\
p_i&=& H(r)^{-1/2}\frac{dx^i}{d\lambda}\,,\nonumber\\
P_r&=&H(r)^{1/2}\frac{dr}{d\lambda}\,,\nonumber\\
J&=&H(r)^{1/2}r^2\frac{d\phi}{d\lambda}\,,
\eea
where $\tau=\mu\lambda $, with $\lambda$ an affine parameter and $\tau$ is the proper time. The mass-shell condition in Hamiltonian form is
$\mathcal{H}=-\mu^2$, 
and can be cast in the form  
\bea
P_r^2&=&\mathcal{E}^2H(r)-\frac{J^2}{r^2}-\mu^2H(r)^{1/2}\,,\nonumber\\
\mathcal{E}^2&=&E^2-\bf{p}^2\,,
\eea
or equivalently
\bea\label{drdtau1}
\left(\frac{dr}{d\lambda}\right)^2&=&\mathcal{E}^2-\frac{J^2}{r^2H(r)}-\frac{\mu^2}{H(r)^{1/2}}\nonumber\\
&=& \mathcal{E}^2
-\frac{J^2r^2}{(r^4+L^4)}-\frac{r^2\mu^2}{\sqrt{r^4+L^4}}\,.
\eea

Passing to the proper-time parametrization the radial equation can be written in the form
\beq
\left(\frac{dr}{d\tau}\right)^2
=
\hat {\mathcal E}^2
-
V_{\rm eff}(r),
\eeq
where the effective potential reads
\beq
V_{\rm eff}(r)
=
\frac{1}{\sqrt{H(r)}}
+
\frac{\hat{J}^2L^2}{r^2H(r)}\,.
\eeq
where we used the notation
\bea
\hat {\mathcal E}=\frac{\mathcal E}{\mu}\,,\quad \hat J=\frac{J}{\mu L}\,.
\eea
\subsection{Circular geodesics}
Circular orbits are characterized by
\beq
\dot r=0,
\qquad
\frac{dV_{\rm eff}}{dr}=0\,,
\eeq
which are solved for the \lq\lq circular values" $r=r_c$, $\hat {\mathcal E}_c=\hat {\mathcal E}|_{r=r_c}$, $\hat J_c=\hat J|_{r=r_c}$ and $H_c=H(r)|_{r=r_c}$.
In addition,
\beq
\frac{d^2V_{\rm eff}}{dr^2}\Bigg|_{r_c}<0\,,\qquad \forall\quad r_c>L\,,
\eeq
so that these orbits are unstable.

\bea\label{critreg}
\frac{r_c}{L}&{=}& (H_c-1)^{-1/4}\,,\nonumber\\ 
\hat{\mathcal{E}}_c&=&\frac{1}{\sqrt{2-H_c}H_c^\frac{1}{4}}\,,\nonumber\\
\hat J_c &=& 
\frac{H_c^{1/4}(H_c-1)^{1/4}}{\sqrt{2-H_c}}\,.
\eea

\begin{figure}
    \centering
    \includegraphics[scale=0.9]{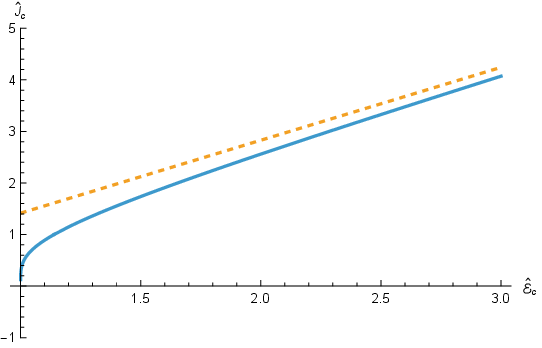}
    \caption{\label{fig:3} The behavior of  $\hat{J}_c$ vs $\hat{\mathcal{E}}_c$, carrying a gauge-invariant information, is shown assuming $L=1$. The dashed curve represents the asymptote $\hat{J}_c=\sqrt{2}\hat{\mathcal{E}}_c$ (actually passing through the origin, even if not deducible from the plot where the horizontal axis starts at $\hat{\mathcal{E}}_c=1)$.}
\end{figure}
Fig. \ref{fig:3} shows the (gauge-invariant) behavior of  $\hat{J}_c$ vs $\hat{\mathcal{E}}_c$.
The corresponding ten velocity  vector $U^\alpha=dx^\alpha/d\tau$  reads
\bea
U&=&\frac{dt}{d\tau}\left(\partial_t +\frac{d\phi}{dt} \partial_\phi\right)= \Gamma\left(\partial_t +\Omega\partial_\phi \right)\,,
\eea
with $r_c>L$  (namely, $1\le H_c< 2$, provided that $r_c>L$) and 
\bea\label{gammaomega}
 \Gamma&=& \frac{H_c^\frac{1}{4}}{\sqrt{2-H_c}}\,,\nonumber\\ 
\Omega &=& 
\frac{1}{L}\frac{(H_c-1)^{3/4}}{\sqrt{H_c}}\,.
\eea
A natural variable associated with the (gauge-invariant) frequency is the following
\beq
\label{y_Omega}
y_{\Omega}=(L\Omega)^{4/3}=\frac{H_c-1}{H_c^{2/3}}=\frac{u^4}{(1+u^4)^{2/3}}
\eeq
where $u=L/r_c$.

\subsection{Radial infall orbits}

Let us assume that all the compact momenta along the internal directions to be vanishing $\vec{p}=0$ \footnote{In fact, such momenta only contribute as an effective mass.}. 
The condition for time-like geodesic in radial fall is
\beq
-1=-\frac{\dot{t}^2}{\sqrt{H(r)}}+\dot{r}^2\sqrt{H(r)}\,,
\eeq
which, using \eqref{canmom}, can be rewritten as
\beq
\frac{dr}{dt}=\frac{\sqrt{\hat {\mathcal E}^2-\frac{1}{\sqrt{H(r)}}}}{\hat {\mathcal E}\sqrt{H(r)}}\,.
\eeq
The previous Eqs. \eqref{canmom} can be integrated exactly assuming the particle at rest at infinity ($\hat {\mathcal E}=1$), leading to
\begin{widetext}
            \beq
t(r)=\frac{r^3\left(\sqrt{H}+1\right)^{3/2}}{3 L^2}\Bigg[\frac{\left(4-3 \sqrt{H}\right)
   \sqrt[4]{H}}{\sqrt{H}+1}-\frac{3 \mathcal{H}_1 L^4}{r^4
   \left(\sqrt{H}+
   1\right)^2}+\frac{\mathcal{H}_2 L^8}{r^8
   \left(\sqrt{H}+
   1\right)^4}\Bigg]\,,
    \eeq
    \end{widetext}
    where
    \bea
\mathcal{H}_1&=&\,
   _2F_1\left(\frac{1}{8},\frac{1}
   {2};\frac{9}{8};\frac{L^8}{r^8}
   \left(\sqrt{H}+
   1)\right)^{-4}\right),\nonumber\\
\mathcal{H}_2&=&\,
   {}_2F_1\left(\frac{1}{2},\frac{5}
   {8};\frac{13}{8};\frac{L^8}{r^8}
   \left(\sqrt{H}+
   1\right)^{-4}\right)\,.
    \eea
    Note that the above expression becomes more compact if one introduces the new variable  (suggested by the arguments of the hypergeometric functions)
    \beq
\xi=\sqrt{H}+1\,,
    \eeq
    namely
    \bea
\frac{t(r)}{L}&=&\frac{ \xi^{3/2}(\xi^2-2\xi)^{-3/4}}{3 }\Bigg[\frac{\left(7-3\xi\right)
   \sqrt{\xi-1}}{\xi}\nonumber\\
   &-&\frac{3(\xi^2-2\xi) }{
   \xi^2} \mathcal{H}_1+\frac{(\xi^2-2\xi)^2}{ 
   \xi^4}\mathcal{H}_2\Bigg]\,.
   \eea 

In expanded form, we find
\bea
\frac{t}{L}&=&\frac{\sqrt{2}}{3}\left(\frac{r}{L}\right)^3-\frac{7}{4 \sqrt{2}}\left(\frac{L}{r}\right)+\frac{1}{64 \sqrt{2}}\left(\frac{L}{r}\right)^5\nonumber\\
&-&\frac{19}{4608 \sqrt{2}} \left(\frac{L}{r}\right)^9+\frac{365 }{212992 \sqrt{2}
   }\left(\frac{L}{r}\right)^{13}\nonumber\\
   &-&\frac{1977 }{2228224
   \sqrt{2} }\left(\frac{L}{r}\right)^{17}
   +\frac{23057}{44040192
   \sqrt{2}}\left(\frac{L}{r}\right)^{21}\nonumber\\
   &+&O(L^{25})\,,
\eea
which can be  inverted perturbatively as
\bea
\frac{r_{\rm orb}(t)}{L}&=&\frac{\sqrt[3]{3} 
   }{\sqrt[6]{2}}\left(\frac{t}{L}\right)^\frac13+\frac{7}{12 \sqrt{2}}\frac{L}{t}-\frac{197 }{288\ 2^{5/6}
   \sqrt[3]{3} }\left(\frac{L}{t}\right)^\frac{7}{3}\nonumber\\
   &+&\frac{41683 }{62208
   \sqrt[6]{2} 3^{2/3} }\left(\frac{L}{t}\right)^\frac{11}{3}-\frac{20668027 }{38817792
   \sqrt{2} }\left(\frac{L}{t}\right)^5\nonumber\\
   &+&\frac{33489665549
   }{23756488704\ 2^{5/6}
   \sqrt[3]{3} }\left(\frac{L}{t}\right)^\frac{19}{3}\nonumber\\
   &{-}&\frac{143148954782411
   }{71839621840896
   \sqrt[6]{2} 3^{2/3} }\left(\frac{L}{t}\right)^\frac{23}{3}+O(L^{9})\,.\nonumber\\
\eea
\subsection{Hyperbolic-like orbits}
The scattering angle follows from \eqref{drdtau1} and \eqref{canmom}
\bea\label{scattang}
\Delta\Phi&=&\int_\infty^{r_1}\frac{d\phi}{dr}\nonumber\\
&=&\int_\infty^{r_1}\frac{bdr}{\sqrt{r^4H(r)-b^2r^2-r^4\rho^2\sqrt{H(r)}}}\,,
\eea
where 
\beq
b=\frac{J}{\mathcal{E}},\quad  \rho^{-1}=\frac{\mathcal{E}}{\mu}=\hat {\mathcal E}\,,
\eeq
and where $r_1$ is the biggest root of the denominator of \eqref{scattang} a.k.a. the turning point, whose explicit expression in PM sense is
\begin{widetext}
\bea
r_1&=&\frac{b}{\sqrt{1-\rho^2}}+\frac{L^4\sqrt{1-\rho^2}(\rho^2-2)}{4b^3}+\frac{L^8\sqrt{1-\rho^2}(3\rho^6-21\rho^4+38\rho^2-20)}{32b^7}\nonumber\\
&+&\frac{L^{12}(1-\rho^2)^\frac{5}{2}(7\rho^6-80\rho^4+220\rho^2-168)}{128b^{11}}\nonumber\\
&-&\frac{L^{16}(1-\rho^2)^\frac{7}{2}(77\rho^8-1388\rho^6+6692\rho^4-11816\rho^2+6864)}{2048b^{15}}\nonumber\\
&+&\frac{L^{20}(1-\rho^2)^\frac{9}{2}(231\rho^{10}-5922\rho^8+43036\rho^6-128016\rho^4+166032\rho^2-77792)}{8192b^{19}}+O\left(\frac{1}{b^{23}}\right)\,.
\eea
The scattering angle is given by
\bea
\frac{\chi}{\pi}&=&\frac{2\Delta\Phi-\pi}{\pi}\nonumber\\
&=&\sum_{k=1}^5 \widetilde a_{4k} \left(\frac{L}{b}\right)^{4k}+O(L^{24})\nonumber\\
&=&\frac{3 L^4 \left(\rho ^2-2\right)
   \left(\rho ^2-1\right)}{8 b^4}+\frac{35 L^8 \left(\rho
   ^2-1\right)^2 \left(2 \rho
   ^4-11 \rho ^2+12\right)}{256
   b^8}\nonumber\\
   &+&\frac{231 L^{12} \left(\rho
   ^2-1\right)^3 \left(8 \rho
   ^6-81 \rho ^4+213 \rho
   ^2-160\right)}{8192 b^{12}}\nonumber\\
   &+&\frac{6435 L^{16} \left(\rho
   ^2-1\right)^4 \left(16 \rho
   ^8-251 \rho ^6+1140 \rho ^4-1955
   \rho ^2+1120\right)}{524288
   b^{16}}\nonumber\\
   &+&\frac{46189 L^{20} \left(\rho
   ^2-1\right)^5 \left(128 \rho
   ^{10}-2825 \rho ^8+19155 \rho
   ^6-54875 \rho ^4+69665 \rho
   ^2-32256\right)}{33554432
   b^{20}}+O(L^{24})\,.
   \eea
  Note that in the massless case, $\rho=0$, the scattering angle can be resummed  to give the known result \cite{Bianchi:2021yqs}
   \beq
\frac{\chi_{0}}{\pi}=\frac{\sqrt{2}b}{\sqrt{b^2+\sqrt{b^4-4L^4}}}\,
{}_2F_1\left(
\frac{1}{2},\frac{1}{2};1;
\frac{b^2-\sqrt{b^4-4L^4}}
{b^2+\sqrt{b^4-4L^4}}
\right)-1\,,
   \eeq
   \end{widetext}
   with
   \beq
   \sqrt{b^2\pm \sqrt{b^4-4L^4}}=\sqrt{\frac{b^2+2L^2}{2}}\pm \sqrt{\frac{b^2-2L^2}{2}}\,.
   \eeq

\section{The Couch-Torrence symmetry}

Let's recall some known results concerning D3 brane spacetimes \cite{Bianchi:2021yqs,Bianchi:2022wku}. For massless geodesics in D3 brane geometry the critical regime corresponding to a circular orbit representing the separatrix between scattering and capture is identified by \eqref{critreg}
\beq
r_c=L,\quad  b_c=\sqrt{2}L\,.
\eeq
The CT inversion symmetry maps the horizon $r=0$ to infinity keeping fixed the radius of the photon-sphere $r\to L^2/r$. Under this transformation the metric is invariant modulo an overall conformal factor $ds^2\underset{r\to L^2/r}{\to}L^2/r^2 ds^2\equiv {\mathcal W}^2 ds^2$. Let us mention in passing that with a conformal factor
\beq
{\mathcal W}^2=\frac{L^2}{r^2}\,,
\eeq
in the full ten dimensional metric one has
\beq
\Box \ln {\mathcal W}=-\frac{4}{r^2 \sqrt{H}}\,,\quad (\nabla \ln {\mathcal W})^2=\frac{1}{r^2 \sqrt{H}}\,,
\eeq
so that the following relation holds
\beq
\Box \ln {\mathcal W}+\frac{d-2}{2} (\nabla \ln {\mathcal W})^2=0\,,
\eeq
with $d=10$ and $(d-2)/2=4$. Since  the original 10-dimensional metric has a vanishing Ricci scalar ($R=0$), this implies vanishing of the Ricci scalar also for the conformally rescaled metric, according to the general relation
\bea
  \tilde R &=& {\mathcal W}^{-2}\Big[ R - 2(d-1)\,\Box \ln {\mathcal W} \nonumber\\
&&  - (d-2)(d-1)\, g^{\mu\nu}\,\nabla_\mu \ln {\mathcal W}\, \nabla_\nu \ln {\mathcal W} \Big]\,,
\eea
that is
\beq
 \tilde R = {\mathcal W}^{-2}  R=0\,.
\eeq
Consequently, the massless scalar field equation $\Box\Phi=0$ is also conformally invariant (the scalar field itself scaling as $\Phi\to {\mathcal W}^{-4}\Phi$).

For massless geodesics the effective potential can be written as 
\bea
P_r^2&=&Q_r=\frac{E^2(r^4-b^2r^2+L^4)}{r^4}\nonumber\\
&=&E^2\frac{(r^2-r_1^2)(r^2-r_2^2)}{r^4}\,,
\eea
where the turning points are
\bea
r_1&=&-r_4=\frac{\sqrt{b^2+\sqrt{b^4-4L^4}}}{\sqrt{2}}\,,\nonumber\\
r_2&=&-r_3=\frac{\sqrt{b^2-\sqrt{b^4-4L^4}}}{\sqrt{2}}\,.
\eea
Notice that a relevant property of the turning points is
\beq
r_1 r_2=r_3r_4=L^2=r_c^2\,.
\eeq
The general expression of the scattering angle is
\beq
\Delta\Phi(A,B)=\int_A^B\frac{b dr}{\sqrt{r^4-b^2r^2+L^4}}\,.
\eeq
As a consequence of the CT symmetry, the scattering angle and the falling angle\footnote{Given an impact parameter beyond the critical one, the massless geodesic motion admits two real turning points. As in \cite{Bianchi:2021mft}, we will call \lq\lq scattering angle" the one spanned from infinity to the external turning point and back again to infinity, and \lq\lq falling angle" the one from the internal turning point to the horizon.} are equal
\bea
&&\Delta\Phi(\infty,r_1)=\Delta\Phi(r_2,0)\nonumber\\
&=&\frac{\sqrt{2}b}{\sqrt{b^2+\sqrt{b^4-4L^4}}}K\Bigg[\frac{b^2-\sqrt{b^4-4L^4}}{b^2+\sqrt{b^4-4L^4}}\Bigg]\,,\nonumber\\
\eea
where $K$ is the complete elliptic integral of the first kind
\bea
K(k)&=&F\left(\frac{\pi}{2},k\right)=\int_0^\frac{\pi}{2}\frac{d\theta}{\sqrt{1-k^2\sin^2\theta}}\nonumber\\
&=&\frac{\pi}{2}{}_2F_1\Big[\frac12,\frac12,1,k\Big]\,.
\eea

\section{Scalar wave equation in a D3 brane background}

In the metric \eqref{metr}  
the (source free) wave equation for a massless scalar field reads
\beq
\Box\Phi=g^{-\frac{1}{2}}\partial_M\left(g^\frac{1}{2}g^{MN}\partial_N\right)\Phi=0\,.
\eeq
Using the ansatz
\beq
\Phi=e^{-i\omega t+i{\bf k}\cdot {\bf x}}r^{-\frac{5}{2}}\psi(r)Y_{\ell \vec{m}}(\vec{\theta})\,,
\eeq
with $\ell\, \vec{m}=\ell\, \ell_2\ell_3\ell_4 m$ and
\beq
\label{lap_arm}
\nabla_{S^5}Y_{\ell \vec{m}} =-\ell(\ell+4)Y_{\ell \vec{m}}\,,
\eeq
and assuming
\beq
k^i=0\,,
\eeq
without loss of generality since such a presence of the compact momenta would represent an effective Kaluza-Klein mass, one finds the radial equation in canonical form without any loss of generality~\footnote{A nonzero $k^2$ is equivalent to a redefinition of $\omega$ according to $\omega^2-k^2=\tilde \omega^2$.}, 
\beq\label{qofr}
\psi''(r)+Q(r)\psi(r)=0\,,
\eeq
with
\beq
Q(r)=\frac{4\omega^2(r^4+L^4)-r^2(4\ell(\ell+4)+15)}{4r^4}\,,
\eeq
while in terms of the rescaled wave function $R(r)$
\beq
R(r)=r^{-\frac{5}{2}}\psi(r)
\eeq
the radial equation becomes\footnote{Note that the zero-frequency limit of this equation cancels out $L$, and hence it corresponds to the trivial flat spacetime case, with two independent solutions $r^\ell$ (regular at the horizon), $r^{-\ell-4}$ (regular at infinity).}
\beq\label{qofr2}
R''(r){+}\frac{5R'(r)}{r}{+}\left[\left(1{+}\frac{L^4}{r^4}\right)\omega^2{-}\frac{\ell(\ell{+}4)}{r^2}\right]R(r){=}0\,.
\eeq
Explicitly, for the solution of the angular part \eqref{lap_arm} involves the spherical harmonics of $S_5$, i.e., 
\bea
\begin{aligned}
\nabla^2_{S^5}
={}&
\frac{1}{\sin^4\theta_1}
\frac{\partial}{\partial\theta_1}
\left(
\sin^4\theta_1
\frac{\partial}{\partial\theta_1}
\right)
\\[2mm]
&+
\frac{1}{\sin^2\theta_1}
\frac{1}{\sin^3\theta_2}
\frac{\partial}{\partial\theta_2}
\left(
\sin^3\theta_2
\frac{\partial}{\partial\theta_2}
\right)
\\[2mm]
&+
\frac{1}{\sin^2\theta_1\sin^2\theta_2}
\frac{1}{\sin^2\theta_3}
\frac{\partial}{\partial\theta_3}
\left(
\sin^2\theta_3
\frac{\partial}{\partial\theta_3}
\right)
\\[2mm]
&+
\frac{1}{\sin^2\theta_1\sin^2\theta_2\sin^2\theta_3}
\frac{1}{\sin\theta_4}
\frac{\partial}{\partial\theta_4}
\left(
\sin\theta_4
\frac{\partial}{\partial\theta_4}
\right)
\\[2mm]
&+
\frac{1}{
\sin^2\theta_1\sin^2\theta_2
\sin^2\theta_3\sin^2\theta_4
}
\frac{\partial^2}{\partial\theta_5^2}\,,
\end{aligned}
\eea
so that from Eq. \eqref{lap_arm} one finds for $Y_{\ell \vec{m}}\equiv Y_{\ell\,\ell_2 \ell_3 \ell_4 m}(\theta_1,\ldots,\theta_5)$
\beq
\begin{aligned}\label{YPL}
Y_{\ell\,\ell_2 \ell_3 \ell_4 m}(\theta_1,\ldots,\theta_5)
={}&
N_{\ell\,\ell_2 \ell_3 \ell_4 m}
\,e^{i m\theta_5}\\
&\times (\sin\theta_4)^{|m|}
C_{\ell_4-|m|}^{|m|+\frac{1}{2}}(\cos\theta_4)
\\
&\times
(\sin\theta_3)^{\ell_4}
C_{\ell_3-\ell_4}^{\ell_4+1}(\cos\theta_3)
\\
&\times
(\sin\theta_2)^{\ell_3}
C_{\ell_2-\ell_3}^{\ell_3+\frac{3}{2}}(\cos\theta_2)
\\
&\times
(\sin\theta_1)^{\ell_2}
C_{\ell-\ell_2}^{\ell_2+2}(\cos\theta_1).
\end{aligned}
\eeq
where the functions $C_n^\lambda(x)$ denote the Gegenbauer polynomials, and the indices are such that
\beq
\ell\geq \ell_2\geq \ell_3\geq \ell_4\geq |m|\,.
\eeq
In this case, the $\vec m$   notation stands for the following list of quantum numbers:  $\vec{m}=(\ell_2,\ell_3,\ell_4,m)$. For example, $Y_{\ell\,\ell_2 \ell_3 \ell_4 m}(\theta_1,\ldots,\theta_5)=Y_{\ell\,\vec m}(\theta_1,\ldots,\theta_5)$.
Finally, $N$ is a normalization constant. Some properties of the Gegenbauer polynomials are collected in Appendix \ref{appa}.

As an example for $\ell=0$ the allowed values for $\ell_i$ ($i=2,3,4$) and $m$ are such that
\beq
\ell_2=\ell_3=\ell_4=m=0\,;
\eeq
for $\ell=1$
\bea
&&\ell_2=\ell_3=\ell_4=|m|=1\,,\nonumber\\
&&\ell_2=\ell_3=\ell_4=1,\quad m=0\,,\nonumber\\
&&\ell_2=\ell_3=1\quad \ell_4=m=0\,,\nonumber\\
&&\ell_2=1\quad \ell_3=\ell_4=m=0\,,\nonumber\\
&&\ell_2=\ell_3=\ell_4=m=0\,.
\eea
For $\ell=2$ straightforwardly the choices are 20, for $\ell=3$ are 50 and for $\ell=4$ are 105, etc. Given $\ell$, the degeneracy $g_\ell$ is then given by
\beq
g_\ell =\frac{(\ell+1)(\ell+2)^2(\ell+3)}{12}\,.
\eeq
A special form, convenient for the D3 brane spacetimes, consists in using the irreducible, symmetric and traceless representation of $SO(6)$. In this case the harmonics of order $l$ can be represented as
\beq
Y^{I_l}(\Omega_5)=C^{I_l}_{A_1\ldots A_l}n^{A_1}\ldots n^{A_l}\,,
\eeq
where $A=1\ldots 6$, $n_A n_A=1$, and, for example
\bea
n^1&=& c_1\,,\quad  
n^2= s_1c_2\,,\quad
n^3= s_1s_2c_3\,,\nonumber\\
n^4&=& s_1s_2s_3c_4\,,\quad
n^5= s_1s_2s_3s_4c_5\,,\nonumber\\
n^6&=& s_1s_2s_3s_4s_5\,,
\eea
where we used the compact notation $[s_i,c_i]=[\sin \theta_i,\cos\theta_i]$.
For $l=0$ the only harmonic is
\beq
Y^{(0)}=\frac{1}{\pi^{3/2}}\,.
\eeq
For $l=1$ one has six harmonics
\beq
Y^{(1)}=\sqrt{\frac{6}{\pi^3}}n^A,\quad A=1\ldots 6\,.
\eeq
For $l=2$ one has 20 independent harmonics
\beq
Y^{(2)}=N_2 \left(n^A n^B-\frac16 \delta^{AB}\right)\,,
\eeq
etc.

The radial potential in the new coordinate  $y$ is
\bea
Q(y)=Q(r(y))(r'(y))^2+\mathcal{S}(r(y))\,,
\eea
where 
\beq
\mathcal{S}(r(y))=\frac{r'''(y)}{2r'(y)}-\frac{3}{4}\left(\frac{r''(y)}{r'(y)}\right)^2
\eeq
is the Schwarzian derivative relative to the transformation $r=r(y)$.

Eq. \eqref{qofr} can be mapped into the quantum Seiberg-Witten curve (Doubly Reduced Doubly Confluent Heun equation, DRDCHE) with no mass hypermultiplet (i.e., pure gauge)
\bea\label{SW00}
Q_{0,0}(y)=\frac{4qy^2+y(1-4u)+4}{4y^3}\,,
\eea
through the  dictionary 
\bea
u=\frac{(\ell+2)^2}{4},\quad q=\frac{L^4\omega^4}{16},\quad \frac{r}{L}=\frac{\omega L}{2}\sqrt{y}\,,
\eea
and at the same time \eqref{qofr} can be  mapped into the modified Mathieu equation
\beq
\psi''(z)+(\alpha-2\beta\cos(2z))\psi(z)=0\,,
\eeq
through the dictionary
\beq
r=Le^{iz},\quad\alpha=(\ell+2)^2,\quad \beta=4\sqrt{q}\,,
\eeq
with $\psi(z)=r^{-1/2}\psi(r)$. 

In terms of the following (new) coordinate $x$
\beq
r=L\left(\sqrt{x}+\sqrt{1+x}\right)^2\,,
\eeq
the differential equation in canonical form exhibits an effective potential 
\beq
\label{qofx}
Q(x){=}\frac{4 x (x{+}1) \left(8 L^2 (8 x (x{+}1){+}1)
   \omega ^2{-}4 \ell
   (\ell{+}4){-}15\right){+}3}{16 x^2
   (x{+}1)^2}\,,
\eeq
which can be mapped to a qSW curve with 3 flavors (i.e., a Confluent Heun equation, CHE)
\bea
Q_{1,2}(y)&{=}{-}\frac{q'^2}{4}{+}\frac{1{-}(m_1{-}m_2)^2}{4y^2}{+}\frac{1{-}(m_1{+}m_2)^2}{4(1{+}y)^2}\nonumber\\
&-\frac{m_3q'}{y}+\frac{-1+2(m_1^2+m_2^2+q'(1-m_1-m_2)-2u')}{4y(1+y)}\,,
\eea
through the dictionary
\bea
&q'=8iL\omega,\quad u'=(\ell+2)^2+2iL\omega-2L^2\omega^2\nonumber\\
&m_1=\frac{1}{2},\quad m_2=m_3=0\,.
\eea
This peculiar transformation of variables is ruled by the Couch-Torrence (CT) symmetry \cite{Cvetic:2020kwf,Cvetic:2021lss,Bianchi:2021yqs,Bianchi:2022wku,Akhond:2026suo} mentioned above.
We can now pass on to determine special types of solutions of the radial equation.\\

\section{PN-type solutions to the radial equation}

Eq. \eqref{qofr} can be rewritten as
\bea
\label{eqfin}
r^2\psi''(r){+}\Big[r^2\omega^2{+}\frac{L^4\omega^2}{r^2}-\frac{15}{4}-\ell(\ell+4)\Big]\psi(r){=}0\,.\nonumber\\
\eea
Introducing  the PN weight for $\omega\to \eta \omega$ and expanding in $\eta$ we find the following solutions: $R_{\rm in, PN}$, regular at the horizon $r=0$ and $R_{\rm up, PN}$, regular at infinity (also simply related among them because of the fact that the radial equation depends on $\ell$ via the combination $\ell(\ell+4)$ which is left invariant by the transformation $\ell \to -\ell-4$)
\begin{widetext}
\bea
\label{RinupPN}
R_{\rm in, PN}&=&r^{\ell }+\eta^2\omega ^2\Bigg[\frac{ r^{\ell -2}
   \left(L^4 (\ell +3)-r^4 (\ell
   +1)\right)}{4 \left(\ell ^2+4
   \ell +3\right)}\Bigg]+\eta^4 \omega ^4\Bigg[-\frac{L^4
   r^{\ell } \log
   (r/L)}{4 (\ell +1) (\ell +2)
   (\ell +3)}\nonumber\\
   &+&\frac{r^{\ell -4}
   \left(L^8 (\ell +2)^2
   \left(\ell ^2+7 \ell
   +12\right)+4 L^4 r^4 \ell 
   (\ell +4)+r^8 \ell  (\ell +1)
   (\ell +2)^2\right)}{32 \ell 
   (\ell +1) (\ell +2)^2 (\ell
   +3) (\ell +4)}\Bigg]+O\left(\eta^6\right)\,,\nonumber\\
R_{\rm up, PN}&=&\left.R_{\rm in, PN}\right|_{\ell\to-\ell-4}\,.
\eea
\end{widetext}
A careful inspection of Eq. \eqref{eqfin} shows \lq\lq two symmetry blocks" in the potential
\bea
V(\omega, r, \ell)&=&\omega^2 \left( r^2+\frac{L^4}{r^2}\right)-\frac{15}{4}-\ell(\ell+4)\nonumber\\
&=&\omega^2 \left( r^2+\frac{L^4}{r^2}\right)+\frac14-(\ell+2)^2\,.
\eea
The first block $-\frac{15}{4}-\ell(\ell+4)$ has the mentioned symmetry (i.e., invariance property) under $\ell \to -\ell-4$; the second block, $\omega^2 \left( r^2+\frac{L^4}{r^2}\right)$,  more interestingly displays the CT symmetry, $r\to L^2/r$,  in the sense that it is invariant under it.
A first consequence of this invariance is the following.
Consider the rescaled in and up solutions,
\beq
R_{\rm in,PN}^{\rm resc}=r^{-\ell}R_{\rm in,PN}(r)\,,\quad
R_{\rm up,PN}^{\rm resc}=r^{\ell+4}R_{\rm up,PN}(r)\,,
\eeq
both starting as $1$ plus $\eta$-corrections
\bea
R_{\rm in,PN}^{\rm resc}=1+\omega^2 \eta^2 A_2^{\rm in}(r)+\omega^4 \eta^4 A_4^{\rm in}(r)+\ldots \,,\nonumber\\
R_{\rm up,PN}^{\rm resc}=1+\omega^2 \eta^2 A_2^{\rm up}(r)+\omega^4 \eta^4 A_4^{\rm up}(r)+\ldots 
\eea
A direct computation shows that
\beq
A_{2k}^{\rm in}(r)\bigg|_{r\to L^2/r}=A_{2k}^{\rm up}(r)\,,
\eeq
namely, it shows  the CT symmetry at work.
This nice feature of the equation (and associated PN solutions) was never seen before.

Properly speaking, Eq. \eqref{qofr2} is first transformed by rescaling $R(r)$ into $R_1(r)=r^2R(r)$, and the new equation for $R_1(r)$ is (exactly) left invariant via the CT map.

In the next sections we will provide different types of solutions, characterized by proper boundary conditions at the horizon ($r=0$, purely ingoing) and at infinity ($r\to \infty $, purely outgoing), belonging to the class of MST solutions. To compare the PN solutions with these ones, we often find it convenient to supplement them with an additional homogeneous solution exhibiting the same type of logarithmic terms, namely $\ln(r/L)\to -\ln (L^2\omega/(2r))$ in the in solution and $\ln(r/L)\to\ln (\omega r/2)$ in the up solution.

\section{MST-type solutions to the radial equation}

As for the black hole case we can define solutions purely ingoing at the horizon ($r=0$) and purely outgoing at infinity following the prescriptions of the MST standard procedure. This leads to the following \lq\lq in" and \lq\lq up" solutions.

\subsection{In solution}
Eq. \eqref{eqfin} can be rewritten as
\bea\label{spit}
&&r^2\psi''(r)+\Big[\frac{L^4\omega^2}{r^2}-\nu(\nu+4)-\frac{15}{4}\Big]\psi(r)\nonumber\\
&&=\left(\ell(\ell+4)-\nu(\nu+4)-r^2\omega^2\right)\psi(r)\,,
\eea
where we introduced the renormalized angular momentum (RAM) $\nu$. The solution of the left hand side is
\bea\label{soltemp}
\psi^{\rm in}_\nu(r)&=&\sqrt{r}\Big[c_1\Gamma(-\nu-1)J_{-\nu-2}\left(\frac{L^2\omega}{r}\right)\nonumber\\
&&+c_2 \Gamma(\nu+3)J_{\nu+2}\left(\frac{L^2\omega}{r}\right)\Big]\,.
\eea

The \lq\lq in solution" is characterized by ingoing boundary conditions, which are imposed on \eqref{soltemp} through the following choice of the integration constants
\be
c_1=1\,,\quad c_2=-c_1\frac{e^{-i\pi\nu}\Gamma(-\nu-1)}{\Gamma(\nu+3)}\,.
\ee
The \lq\lq in solution" can be then explicitly written as
\bea
\label{eq:psi_in}
\Psi^{\rm in }_\nu(r)&=&\sum_{n=-\infty}^\infty a_n\sqrt{r}\Big[e^{i\pi(n+\nu)} J_{{-}\nu-n{-}2}\left(\frac{L^2\omega}{r}\right)\nonumber\\
&{-}&J_{\nu+n{+}2}\left(\frac{L^2\omega}{r}\right)\Big]\,.
\eea
In Eq. \eqref{spit} the source on the right hand side 
can be rewritten as
\bea
r^2 \psi^{\rm in }_{\nu+n}(r)&=&A_2\psi^{\rm in }_{\nu+n+2}(r)+A_0\psi^{\rm in }_{\nu+n}(r)\nonumber\\
&+& A_{-2}\psi^{\rm in }_{\nu+n-2}(r)\,,
\eea
where
\bea
A_2&=&\frac{L^4\omega^2}{4(n+\nu+2)(n+\nu+3)}\,,\nonumber\\
A_0&=&\frac{L^4\omega^2}{2(n+\nu+1)(n+\nu+3)}\,,\nonumber\\
A_{-2}&=&\frac{L^4\omega^2}{4(n+\nu+1)(n+\nu+2)}\,.
\eea
Thanks to the previous expression, Eq. \eqref{spit} can be written as the following recursion relation
\bea\label{5rec}
&&0=\psi^{\rm in}_{n+\nu-2}(r)\frac{L^4\omega^4}{4(n+\nu+1)(n+\nu+2)}\nonumber\\
&&+\psi^{\rm in}_{n+\nu+2}(r)\frac{L^4\omega^4}{4(n+\nu+2)(n+\nu+3)}\nonumber\\
&&+\psi^{\rm in}_{n+\nu}(r)\Big[(n+\nu)(n+\nu+4)-\ell(\ell+4)\nonumber\\
&&+\frac{L^4\omega^4}{2(n+\nu+1)(n+\nu+3)}\Big]\,,
\eea
where one  reads the coefficients of the recursion
\bea
\alpha_n&=&\frac{L^4 \omega ^4}{4 (\nu
   +n+3) (\nu +n+4)}\,,\nonumber\\
\gamma_n&=&(n+\nu)(n+\nu+4)-\ell(\ell+4)\nonumber\\
&+&\frac{L^4\omega^4}{2(n+\nu+1)(n+\nu+3)}\,,\nonumber\\
\epsilon_n&=&\frac{L^4 \omega ^4}{4 (\nu
   +n) (\nu +n+1)}\,.
\eea
We will consider this recursion relation in the next subsection below.

\subsection{Recursion relation}

The   recursion \eqref{5rec} can be treated as two independent three terms recursion in the following way
\bea
\alpha^{(e)}_na_{2n+2}+\beta^{(e)}_na_{2n}+\gamma^{(e)}_n a_{2n-2}=0\,,\nn\\
\alpha^{(o)}_na_{2n+1}+\beta^{(o)}_na_{2n-1}+\gamma^{(o)}_n a_{2n-3}=0\,,
\eea
where
\bea
&&\alpha^{(e)}_n=\alpha_{2n},\quad \beta^{(e)}_n=\gamma_{2n},\quad \gamma^{(e)}_n=\epsilon_{2n}\,,\nn\\
&&\alpha^{(o)}_n=\alpha_{2n+1},\quad \beta^{(o)}_n=\gamma_{2n+1},\quad \gamma^{(o)}_n=\epsilon_{2n+1}\,.\nn\\
\eea
Performing the shift $n\to n+1$ in the odd recursion relation, we obtain
\bea
\alpha^{(o)}_{n+1}a_{2n+3}+\beta^{(o)}_{n+1}a_{2n+1}+\gamma^{(o)}_{n+1} a_{2n-1}=0\,.
\eea
Consequently, both the three terms (\lq\lq odd" and \lq\lq even") recursion relations are equivalent (i.e., have the same coefficients) and from now on we will avoid the even-odd extra labels \lq\lq e" and \lq\lq o".

The corresponding continuous fraction is given by
\bea
\beta _0-\frac{\alpha _{-1} \gamma _0}{\beta _{-1}-\frac{\alpha _{-2} \gamma
   _{-1}}{\beta _{-2}-\dots}}-\frac{\alpha _0 \gamma _1}{\beta _1-\frac{\alpha _1
   \gamma _2}{\beta _2-\dots}}=0\,,
\eea
and can be solved perturbatively for small $L\omega$ for $\nu$ \cite{Bini:2025ltr,Bini:2025bll,Bini:2026foz}
\begin{widetext}
\bea\label{nuell}
\nu&=&\ell-\frac{L^4 \omega ^4}{4 (\ell
   +1) (\ell +2) (\ell +3)}-\frac{L^8 \omega ^8 \left(15
   \ell ^4+120 \ell ^3+325 \ell
   ^2+340 \ell +108\right)}{64
   \ell  (\ell +1)^3 (\ell +2)^3
   (\ell +3)^3 (\ell +4)}\nonumber\\
   &{-}&\frac{L^{12} \omega ^{12}
   \left(105 \ell ^{10}{+}2100
   \ell ^9{+}17745 \ell ^8{+}82320
   \ell ^7{+}227255 \ell ^6{+}375060
   \ell ^5{+}341795 \ell ^4{+}115800
   \ell ^3{-}55828 \ell ^2{-}56592
   \ell {-}12960\right)}{256 (\ell
   -1) \ell ^2 (\ell +1)^5 (\ell
   +2)^5 (\ell +3)^5 (\ell +4)^2
   (\ell +5)}\nonumber\\
   &-&\frac{5 L^{16} \omega ^{16}}{16384 (\ell
   -2) (\ell -1)^2 \ell ^3 (\ell
   +1)^7 (\ell +2)^7 (\ell +3)^7
   (\ell +4)^3 (\ell +5)^2 (\ell
   +6)}\Big[
   3003 \ell ^{18}+108108
   \ell ^{17}+1738737 \ell
   ^{16}\nonumber\\
   &+&16432416 \ell
   ^{15}+100702602 \ell
   ^{14}+414197784 \ell
   ^{13}+1126933522 \ell
   ^{12}+1816591920 \ell
   ^{11}+755719679 \ell
   ^{10}-3748116372 \ell
   ^9\nonumber\\
   &-&9434293635 \ell
   ^8-9995579568 \ell
   ^7-3433822324 \ell
   ^6+3181653264 \ell
   ^5+3536165488 \ell
   ^4+487183488 \ell
   ^3-919760832 \ell
   ^2\nonumber\\
   &-&534055680 \ell
   -93312000\Big]+O\Big[(L\omega)^{20}\Big]\,.
\eea

For the first values of $\ell$ (for a specific choice of $\ell$ from the beginning, i.e., not using Eq. \eqref{nuell} which, however, at the order $(L\omega)^{4k}$ holds only for $\ell \ge k$) we find
\bea
\nu^{\ell=0}&=&-\frac{i\sqrt{5}}{48}L^4\omega^4+\frac{7i}{27648\sqrt{5}}L^8\omega^8+\frac{11851i}{127401984\sqrt{5}}L^{12}\omega^{12}+O(L^{16}\omega^{16})\,,\nonumber\\
\nu^{\ell=1}&=&1-\frac{L^4\omega^4}{96}+\frac{133L^8\omega^8}{1105920}+\frac{311L^{12}\omega^{12}}{6370099200}+O(L^{16}\omega^{16})\,,\nonumber\\
\nu^{\ell=2}&=&2-\frac{L^4\omega^4}{240}-\frac{137L^8\omega^8}{6912000}+\frac{305843L^{12}\omega^{12}}{2786918400000}+O(L^{16}\omega^{16})\,,\nonumber\\
\nu^{\ell=3}&=&3-\frac{L^4\omega^4}{480}-\frac{709L^8\omega^8}{193536000}-\frac{547681L^{12}\omega^{12}}{39016857600000}+O(L^{16}\omega^{16})\,,
\eea
where $\nu^{\ell=0}$ is purely imaginary, as already noted in Ref.  \cite{Gubser:1998iu}.
\end{widetext}
In the eikonal (large $\ell$) limit \eqref{nuell} resums in terms of a hypergeometric function \cite{Lerche:1996xu,Bini:2026uqw}, namely up to the Next-to-Leading order
\beq
\nu=\nu^{\rm LO}+\nu^{\rm NLO}\,,
\eeq
with
\bea
\nu^{\rm LO} &=& \ell-\frac{L^4 \omega ^4}{4 \ell ^3}-\frac{15 L^8
   \omega ^8}{64 \ell ^7}-\frac{105 L^{12} \omega ^{12}}{256 \ell ^{11}}\nonumber\\
&-&\frac{15015 L^{16} \omega ^{16}}{16384
   \ell ^{15}}-\frac{153153 L^{20}
   \omega ^{20}}{65536 \ell ^{19}} \nonumber\\
&-&\frac{6789783 L^{24} \omega ^{24}}{1048576 \ell ^{23}}+O(L^{28}\omega^{28})\,,\nonumber\\
\nu^{\rm NLO} &=& \frac{3 L^4 \omega ^4}{2 \ell ^4}+\frac{105 L^8
   \omega ^8}{32 \ell ^8} +\frac{1155 L^{12} \omega ^{12}}{128 \ell ^{12}}\nonumber\\
&+&\frac{225225 L^{16} \omega ^{16}}{8192
   \ell ^{16}}
+\frac{2909907 L^{20}
   \omega ^{20}}{32768 \ell ^{20}}\nonumber\\
&+& \frac{156165009 L^{24} \omega ^{24}}{524288 \ell ^{24}}+O(L^{28}\omega^{28})\,,
\eea
which resum as
\bea
\nu&=&\ell\,{}_2F_1\Big[-\frac{1}{4},\frac{1}{4},1,\frac{4L^4\omega^4}{\ell^4}\Big]\nonumber\\
&{+}&\frac{2}{L^4\omega^4}\Bigg\{\frac{\ell}{\sqrt{\ell^2{+}2L^2\omega^2}}{}_2F_1\Big[\frac12,\frac12,1,\frac{4L^2\omega^2}{\ell^2{+}2L^2\omega^2}\Big]{-}1\Bigg\}\,.\nonumber\\
\eea
Equivalently both these terms can be represented in terms of the more compact expression
\bea
\nu^{\rm LO}+\nu^{\rm NLO}=(\ell+2){}_2F_1\Big[-\frac14,\frac14,1,\frac{4L^4\omega^4}{(\ell+2)^4}\Big]-2\,,\qquad
\eea
with an expansion properly performed to include only LO and NLO temrs.

Note that in the $R_{\rm in, up}$ PN solutions, Eq. \eqref{RinupPN} one can resum the logarithms according to the relation
\bea
R_{\rm in}(r)&=&e^{(\ell -\nu)\ln \left(\frac{L^2\omega}{2r}\right)}\widetilde R_{\rm in}(r)\nonumber\\
&=& \left(\frac{L^2\omega}{2r}\right)^{(\ell -\nu)}\widetilde R_{\rm in}(r)\,,
\eea
where 
\beq
\widetilde R_{\rm in}(r)=R_{\rm in}(r)\bigg|_{\ln \to 0}\,.
\eeq
In other words, the PN solution \lq\lq knows" the renormalized angular momentum $\nu$, as was already noted in Eq. (4.14) of Ref. \cite{Bini:2026foz}.

\subsection{Up solution}

The \lq\lq up" solution is found imposing outgoing boundary conditions on \eqref{eqfin}. To start, let's rewrite such equation as
\bea
&&r^2\psi''(r)+\Big[r^2\omega^2-\nu(\nu+4)-\frac{15}{4}\Big]\psi(r)\nonumber\\
&&=\Big[\ell(\ell+4)-\nu(\nu+4)-
\frac{L^4\omega^2}{r^2}\Big]\psi(r)\,.
\eea
The left hand side is solved in terms of Bessel functions 
\beq
\psi^{\rm up}_\nu(r)=\sqrt{r}\Big[d_1J_{\nu+2}(\omega r)+d_2Y_{\nu+2}(\omega r)\Big]\,.
\eeq
Imposing the outgoing boundary conditions one finds
\beq
d_2=i d_1,\quad d_1=1\,,
\eeq
and the solution becomes writable in terms of Hankel functions
\beq
\psi^{\rm up}_\nu(r)=\sqrt{r}H^{(1)}_{\nu+2}(\omega r)\,.
\eeq
The full \lq up' solution is then given by
\beq
\label{eq:psi_up}
\Psi^{\rm up}_\nu(r)=\sum_{n=-\infty}^\infty a_n \sqrt{r}H^{(1)}_{\nu+n+2}(\omega r)\,.
\eeq
The source can be rewritten thanks to the following identity
\beq
\frac{\psi^{\rm up}_{\nu+n}(r)}{r^2}=A_2 \psi^{\rm up}_{\nu+n+2}(r)+A_0 \psi^{\rm up}_{\nu+n}(r)+A_{-2} \psi^{\rm up}_{\nu+n-2}(r)\,,
\eeq
where
\bea\label{623}
A_2&=&\frac{\omega^2}{4(n+\nu+2)(n+\nu+3)}\,,\nonumber\\
A_0&=&\frac{\omega^2}{2(n+\nu+1)(n+\nu+3)}\,,\nonumber\\
A_{-2}&=&\frac{\omega^2}{4(n+\nu+2)(n+\nu+1)}\,.
\eea
The recursion relation following Eq. \eqref{623} coincides with \eqref{5rec}. As an example, the PN expansions for $\ell=1$ of both \lq in' and \lq up' solutions are collected in Appendix \ref{appb}.

A final comment is in order. The MST-type up solution satisfies the equation
\bea
&&r^2\psi''(r)+\Big[r^2\omega^2-\nu(\nu+4)-\frac{15}{4}\Big]\psi(r)\nonumber\\
&&=\Big[\ell(\ell+4)-\nu(\nu+4)-
\frac{L^4\omega^2}{r^2}\Big]\psi(r)\,,
\eea
while the MST-type in solution satisfies the equation
\bea
&&r^2\psi''(r)+\Big[\frac{L^4\omega^2}{r^2}-\nu(\nu+4)-\frac{15}{4}\Big]\psi(r)\nonumber\\
&&=\Big[\ell(\ell+4)-\nu(\nu+4)-
r^2\omega^2\Big]\psi(r)\,,
\eea
These two equations have their homogeneous part which has broken the CT symmetry (i.e. only contain  the term $\frac{L^4\omega^2}{r^2}$ or the term $r^2\omega^2$) but one equation goes exactly into the other under the CT symmetry.
This is a big difference   with respect to the homogeneous equation used for the PN solution,
\bea
&&r^2\psi''(r)+\Big[\omega^2\left(r^2+\frac{L^4}{r^2}\right)-\frac{15}{4}-\ell(\ell+4)\Big]\psi(r)=0\,,\qquad
\eea
which instead respects the CT symmetry.
However,  for example in the case $\ell=1$, we find  the following relation
\bea
R_{\rm in, MST}^{\ell=1}&=& C_1 R_{\rm in, PN}^{\ell=1}+C_2 R_{\rm up, PN}^{\ell=1}\nonumber\\
&=& C_1 r A_1 (r)  +C_2 \frac{1}{r^5} A_1\left(\frac{L^2}{r}\right)\,,
\eea
which provides a way to restore this symmetry and $A_1(r)$ (the same function in both terms) is a certain function of $r$. Here $C_1$ and $C_2$ are PN constants needed to express a family of solution into the other.
The same is true for any other $\ell$.

\subsection{Asymptotics and Borel-Laplace resummation}

Having solutions $\Psi^{\rm in}$ and $\Psi^{\rm up}$ given by infinite sums as in~\eqref{eq:psi_in} and~\eqref{eq:psi_up} is a consequence of the singular behaviour of the differential equation in~\eqref{eqfin} at both $r=0$ and $1/r=0$. An alternative approach to solve equation~\eqref{eqfin} is to start with formal solutions and then study their summability properties. In fact, formal solutions of differential equations with a so-called {\em single level-}$s$ {\em irregular singularity}~\cite{diverg-resurg--ii} at $r=0$ admit factorially divergent solutions of the form 
\be\label{eq:formal_sol}
e^{-\lambda r^{-1/s}} r^\tau \sum_{n=0}^\infty c_n r^n\,,
\ee
for some constants $\lambda, \tau\in\mathbb{C}$ and the coefficients $c_n$ satisfying 
\be\label{eq:gevrey-s}
|c_n|\lesssim A^n (n!)^s\,, \quad n\geq 0\,.
\ee
A formal series $\sum_{n\ge 0}c_n r^n$, whose coefficients satisfy~\eqref{eq:gevrey-s}, belongs to the set $\mathbb{C}[\![r]\!]_s$.  
The constant $s$ measures both the growth of the coefficients and the exponential rate. In fact, a formal solution of the form~\eqref{eq:formal_sol} should be regarded as the asymptotic expansion as $r\to0$ of an analytic solution. The existence of such solution is guaranteed by the Main Asymptotic Existence Theorem~\cite[Theorem~3.1]{diverg-resurg-iii}, and in many cases it can be computed by the Borel-Laplace re-summation \cite{FF23}. The latter is a piece-wise linear holomorphic function, whose jumps are described by the Stokes phenomenon. 

Here, we will compute the formal {\em in} and {\em up} solutions and discuss their resummation properties. 

\begin{figure}
\includegraphics[scale=0.5]{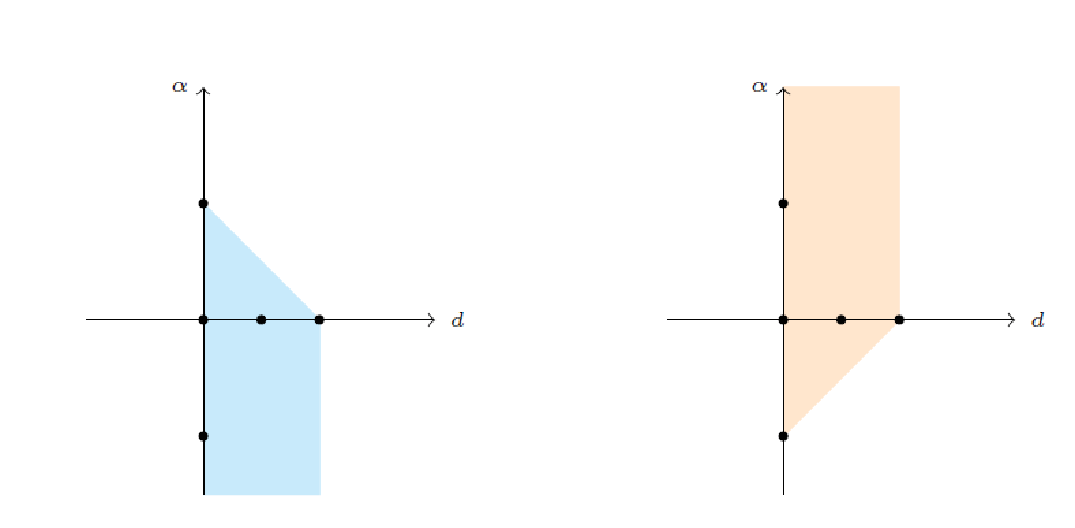} 
\caption{\label{fig:4} The bullet points of coordinates $(d,\alpha)\in\mathbb{N}\times\mathbb{Q}$ correspond to non-zero terms of the form $r^{\alpha}\Theta^d$ in equation~\eqref{eq:spit_Theta} and they constitute the support {\em supp} of the differential operator. In particular, the support for equation~\eqref{eq:spit_Theta} is given by the points $\{(0,0),(1,0),(2,0),(0,-2),(0,2)\}$. Left panel:  The Newton Polygon of the equation~\eqref{eq:spit_Theta} for the singularity $1/r=0$. It is the convex hull of $\{(x, \alpha - y) : (d, \alpha)\in {\rm supp}, 0\le x < d, y \ge 0\}$.  Right panel: The Newton Polygon of the equation~\eqref{eq:spit_Theta} for the singularity $r=0$. It is the convex hull of $\{(x, \alpha + y) : (d, \alpha)\in {\rm supp}, 0\le x < d, y\ge 0\}$. 
In both cases, the polygon presents an edge of slope $\pm 1$, meaning that the points $r=0$ and $1/r=0$ are single level-$1$ irregular singularity. }
\end{figure}

We assume $\omega\neq 0$, indeed when $\omega= 0$ the equation~\eqref{eqfin} admits convergent solution both at $1/r=0$ and $r=0$. First of all, we will compute the constant $s$ in~\eqref{eq:formal_sol} by looking at the slope of the Newton Polygon of~\eqref{eqfin}. Following the convention of~\cite{NP}, let us introduce the log-derivatives $\Theta=r\partial_r$ and rewrite~\eqref{eqfin} as 
\be\label{eq:spit_Theta}
\Theta^2 \psi - \Theta\psi +\Big[r^2\omega^2+\frac{L^4\omega^2}{r^2}-\frac{15}{4}-\ell(\ell+4)\Big]\psi=0\,.
\ee  
Notice that $\Theta$ is invariant (up to a sign) under $r\mapsto1/r$, thus the Newton Polygon can be easily computed in both regimes as illustrated in Fig. \ref{fig:4} (left and right panel), which represents the leading order behaviour of the formal solution at $1/r=0$ and $r=0$, respectively.

Hence, the formal solutions at $1/r=0$ are given by the divergent series $\widetilde{\Phi}_\pm(r)\in e^{\pm i\omega r} \mathbb{C}[\![r^{-1}]\!]_1 $ such that
\be
\widetilde{\Phi}_{\infty,\pm}(r)= e^{\pm i\omega r}(1\pm\frac{c_{1}}{r} + \frac{c_{2}}{r^2}\pm\cdots) \,, 
\ee
and the coefficients $c_n$ satisfy the three-terms recursion
\begin{multline}\label{eq:rec_an}
2i \omega (n + 1) c_{n+1} =  (n (n + 1) -\frac{15}{4}-\ell(\ell+4)) c_n \\
 + \omega^2 L^4 c_{n-2} \,,
\end{multline}
with 
\bea
c_0 & = & 1 \,,\nonumber\\
c_1 &= & i\frac{15+4\ell(\ell+4)}{8\omega}\,,\nonumber\\
c_2 & =& i\frac{7+4\ell(\ell+4)}{16\omega}c_1\,. 
\eea
Similarly, the formal solutions at $r=0$ are given by the divergent series $\widetilde{\Phi}_{0,\pm}(r)\in e^{\pm \frac{i\omega L^2}{r}} r \mathbb{C}[\![r]\!]_1 $ such that
\be
\widetilde{\Phi}_{0,\pm}(r)= e^{\pm \frac{i\omega L^2}{r}}r(1\pm b_{1} r + b_{2}r^2\pm \cdots)\,,
\ee
and the coefficients $b_n$ satisfy the three-terms recursion 
\begin{multline}\label{eq:rec_bn}
2i L^2\omega (n + 1) b_{n+1} =  (n (n + 1) -\frac{15}{4}-\ell(\ell+4)) b_n \\
 + \omega^2 b_{n-2} \,,
\end{multline}
with 
\bea
b_0 & = & 1 \,,\nonumber\\
b_1 &= & i\frac{15+4\ell(\ell+4)}{8L^2\omega}\,,\nonumber\\
b_2 & =& i\frac{7+4\ell(\ell+4)}{16L^2\omega}b_1\,. 
\eea
Equivalently, by comparing the recursions in~\eqref{eq:rec_an} and~\eqref{eq:rec_bn}, we find $b_n L^{2n}=c_n$. This is in agreement with the CT symmetry, which for the solutions gives
\be\label{eq:CT_formal}
\widetilde{\Phi}_{0,\pm}(r)=r\,\widetilde{\Phi}_{\infty,\pm}(L^2/r)\,.
\ee
By the classical theory of Borel-Laplace summability, these divergent asymptotic series $\widetilde{\Phi}_{0,\pm}(r)$ and $\widetilde{\Phi}_{
\infty,\pm}(r)$ can be resummed in any directions $\vartheta\neq\pm\frac{\pi}{2}$. In particular, analytic solutions of~\eqref{eqfin} are given by the resummation of $\widetilde{\Phi}_{\infty,\pm}(r)$ and $\widetilde{\Phi}_{0,\pm}(r)$ using the Laplace kernels $e^{-r\zeta}$ and $e^{-\frac{\zeta}{r}}$, respectively. Explicitly, 
\be
\Phi_{\infty,\pm}(r)=e^{\pm i\omega r}r\int_0^\infty e^{-r\zeta} \varphi_{\pm}(\zeta)\,d\zeta\,,
\ee
where $\varphi_{\pm}$ denotes the sum of the convergent series $\sum_{n\geq 0}(\pm1)^nc_{n}\frac{\zeta^n}{n!}$, and similarly
\be
\Phi_{0,\pm}(r)=e^{\pm i\frac{L^2\omega}{r}}\int_0^\infty e^{-\frac{\zeta}{r}} \varphi_{\pm}(\zeta/L^2)\,d\zeta\,.
\ee
In addition, since we are interested in real solutions and the direction $\vartheta=0$ is admissible, the Stokes data will not play any role in the study of the resummation process. However, we expect that since only two asymptotic behaviours are relevant near $r=0$ (resp. $r=\infty$), the Stokes matrices will be of the form $[1,S; 0,1]$ for some constant $S\in\mathbb{C}$. 

Finally, by looking at the leading order asymptotics of $\Psi^{\rm in}$ and $\Psi^{\rm up}$ in equations~\eqref{eq:psi_in} and~\eqref{eq:psi_up} we expect that they are proportional to the resummed solutions $\Phi_{0,+}(r)$ and $\Phi_{\infty,+}(r)$, respectively. 

\section{Scalar wave equation sourced by a particle moving on a circular orbit}

Consider the massless scalar field equation sourced by the matter-energy distribution associated with a massive test particle moving along an equatorial circular geodesic of the background
\beq
\Box\Phi=-4\pi \rho \,,
\eeq
with the parametric equation of the orbit given by
\beq
x(\tau)=\left(\Gamma\tau,0,0,0,r_c,\frac{\pi}{2},\frac{\pi}{2},\frac{\pi}{2},\frac{\pi}{2},\Omega \Gamma \tau\right)\,.
\eeq
The delta functions corresponding to the three compact directions can be represented as
\beq
\delta(x_i)=\sum_{n_i}\frac{e^{i\frac{n_i}{R_i}x_i}}{2\pi R_i}\,,
\eeq
where the quantized momenta have been represented as $k_i=n_i/R_i$, where $R_i$ are the compactification radii. For the purposes of the present paper we can assume $n_i=0$. Therefore, the new effective scalar charge turns out to be given by
\beq
q_{D3}=\frac{q}{(2\pi)^3R_1R_2R_3}\,.
\eeq
Finally, the scalar source can be written as
\bea
\rho(t,{\mathbf x}, r, \Omega_5)&=&q\int  d\tau \frac{1}{\sqrt{-g}}\delta^{(10)} (x-x(\tau))\nonumber\\
&=& \frac{q_{D3}}{u^t r_c^5 \sqrt{H(r_c)}}\delta(r{-}r_c) \delta_{S_5}(\Omega{-}\Omega')\,,\qquad
\eea
where the completeness relation for the spherical harmonics on $S_5$ reads
\beq
\delta_{S_5}(\Omega-\Omega')=\sum_{\ell, \vec m}  Y_{\ell\vec m}\left(\Omega\right)Y^*_{\ell\vec{m}}(\Omega')\,,
\eeq
having defined
\bea
\Omega&=&\left(\theta_1,\theta_2,\theta_3,\theta_4,\theta_5\right)\,,\nonumber\\
 \Omega'&=&\left(\frac{\pi}{2},\frac{\pi}{2},\frac{\pi}{2},\frac{\pi}{2},\Omega t\right)\,,\nonumber\\
  \widetilde \Omega &=&\left(\frac{\pi}{2},\frac{\pi}{2},\frac{\pi}{2},\frac{\pi}{2},0\right)\,,\nonumber\\
 \vec{m}&=&(\ell_2,\ell_3,\ell_4,m)
\eea
Therefore,
\bea
\rho(t,{\mathbf x}, r, \Omega_5)
&=& \frac{q_{D3}}{u^t r_c^5 \sqrt{H(r_c)}}\delta(r{-}r_c)\times \nonumber\\
&&\sum_{\ell  \vec m \omega}  Y_{\ell\vec m}\left(\Omega\right)Y^*_{\ell\vec{m}}(\frac{\pi}{2},\frac{\pi}{2},\frac{\pi}{2},\frac{\pi}{2},0)e^{-im \Omega t}\,,\nonumber\\
&=& \frac{2\pi q_{D3}}{u^t r_c^5 \sqrt{H(r_c)}}\delta(r{-}r_c)\times \nonumber\\
&&\sum_{\ell  \vec m \omega }  Y_{\ell\vec m}\left(\Omega\right)Y^*_{\ell\vec{m}}( \widetilde \Omega)\int \frac{d\omega}{2\pi}e^{-i\omega t} \delta(\omega-m\Omega)\,.\nonumber\\
\eea
Equivalently
\bea
\rho(t,{\mathbf x}, r, \Omega_5)
&=& \sum_{\ell  \vec m \omega}  Y_{\ell\vec m}\left(\Omega\right)\int \frac{d\omega}{2\pi}e^{-i\omega t} \rho_{\ell  \vec m  \omega}(r)\,,\qquad
\eea
with
\beq
\label{rho_lm_def}
\rho_{\ell  \vec m \omega}=\frac{2\pi q_{D3}}{u^t r_c^5 \sqrt{H(r_c)}}\delta(r{-}r_c)
Y^*_{\ell\vec{m}}( \widetilde \Omega) \delta(\omega-m\Omega)\,.
\eeq

The formulae for the radiated energy can be derived from the associated energy-momentum tensor, which, for a massless complex scalar field, is given by
\beq
8\pi T_{\mu\nu}^{\rm scal}=\partial_\mu\Phi^*\partial_\nu\Phi+\partial_\mu\Phi\partial_\nu\Phi^*-g_{\mu\nu}\partial_\lambda\Phi^*\partial^\lambda\Phi\,,
\eeq
so that
\beq
\frac{d^2E}{dtd\Omega_5}=\lim_{r\to\infty}\left(r^5{T^{\rm scal\,r}}_t\right)\,.
\eeq
Since the D3 brane metric is diagonal, we find
\beq
8\pi T^{\rm scal}_{rt}=\Phi_{,r}^*\Phi_{,t}+\Phi_{,r}\Phi_{,t}^*\,,
\eeq
which implies
\bea
8\pi {T^{\rm scal\,r}}_t&=&g^{rr}(\Phi_{,r}^*\Phi_{,t}+\Phi_{,r}\Phi_{,t}^*)\nonumber\\
&=&H(r)^{-\frac{1}{2}}\Phi_{,r}^*\Phi_{,t}+{\rm c.c.}\,.
\eea
Let us write
\beq\label{phidec}
\Phi=\sum_{\ell \vec{m}}Y_{\ell \vec{m}}\int \frac{d\omega}{2\pi} e^{-i\omega t}R_{\ell  {\vec m}  \omega}(r)\,,
\eeq
where we considered the quantized momenta along the compactified directions to be vanishing.

Explicitly, the radiated energy is given by
\begin{widetext}
    \bea\label{dedt}
\frac{dE}{dt}&=&\lim_{r\to\infty}\int d \Omega_5 r^5 {T^{\rm scal\,r}}_t\nonumber\\
&=&\lim_{r\to\infty}\frac{r^7}{8\pi\sqrt{r^4+L^4}}\sum_{\ell \vec{m}}\int\frac{d\omega}{2\pi}\frac{d\omega'}{2\pi}\Big[(-i\omega)e^{-i(\omega-\omega')t}R(r)\frac{d}{dr}R^*(r)\nonumber\\
&+&(i\omega')e^{i(\omega-\omega')t}R^*(r)\frac{d}{dr}R(r)\Big]\,,
    \eea
    \end{widetext}
where we used the orthogonality relation
    \beq
\int_{S_5}d\Omega_5 Y_{\ell \vec{m}}Y^*_{\ell' \vec{m}'}=\delta_{\ell\ell'}\delta_{\vec{m} \vec{m}'}\,.
    \eeq

The (angles and time decomposed) wave equation in presence of the source becomes
\bea
\Box\Phi&=&\sum_{\ell \vec{m}}Y_{\ell \vec{m}}(\Omega)\int \frac{d\omega}{2\pi} e^{-i\omega t}\mathcal{L}_r R_{\ell\vec{m}\omega}(r)\nonumber\\
&=&-4\pi\sum_{\ell, \vec m}Y_{\ell, \vec m}(\Omega)\int \frac{d\omega}{2\pi} e^{-i\omega t}
\rho_{\ell  \vec m  \omega}(r)\,,
\eea
where
\bea
\mathcal{L}_r&=&\frac{1}{\sqrt{H(r)}}\partial_r^2+\frac{5}{r\sqrt{H(r)}}\partial_r\nonumber\\
&&+\Bigg[\omega^2\sqrt{H(r)}-\frac{\ell(\ell+4)}{r^2\sqrt{H(r)}}\Bigg]\,,
\eea
and $\rho_{\ell  \vec m \omega}$ defined in Eq. \eqref{rho_lm_def}.
Consequently
\beq\label{820}
\mathcal{L}_r\left(R(r)\right)=-4\pi \rho_{\ell  \vec m  \omega}(r)\,.
\eeq
Using the definition of the spherical harmonics on $S_5$ in \eqref{YPL}, we have
\begin{widetext}
\bea
&&Y^{*}_{\ell\vec m}\left(\frac{\pi}{2},\frac{\pi}{2},\frac{\pi}{2},\frac{\pi}{2},\Omega t\right)=P_{\ell \ell_2 \ell_3 \ell_4 m}e^{-im\Omega t}\nonumber\\
&&P_{\ell \ell_2 \ell_3 \ell_4 m}=\frac{2^{\ell +\ell _2+\ell _3+\ell _4+\frac{3}{2}} \sqrt{\ell +2} \sqrt{2
   \ell _2+3} \sqrt{\ell _3+1} \sqrt{2 \ell _4+1}}{\sqrt{\pi }}\nonumber\\
   &&\times\frac{1}{\sqrt{\Gamma \left(\ell {-}\ell _2{+}1\right) \Gamma \left(\ell {+}\ell
   _2{+}4\right) \Gamma \left(\ell _2{-}\ell _3{+}1\right) \Gamma \left(\ell
   _2{+}\ell _3{+}3\right) \Gamma \left(\ell _3{-}\ell _4{+}1\right) \Gamma
   \left(\ell _3{+}\ell _4{+}2\right) \Gamma \left(\ell _4{-}m{+}1\right) \Gamma
   \left(m{+}\ell _4{+}1\right)}}\nonumber\\
   &&\times \frac{\Gamma \left(\frac{1}{2} \left(\ell +\ell _2+4\right)\right) \Gamma
   \left(\frac{1}{2} \left(\ell _2+\ell _3+3\right)\right) \Gamma
   \left(\frac{1}{2} \left(\ell _3+\ell _4+2\right)\right) \Gamma
   \left(\frac{1}{2} \left(m+\ell _4+1\right)\right)}{\Gamma
   \left(\frac{1}{2} \left(-\ell +\ell _2+1\right)\right) \Gamma
   \left(\frac{1}{2} \left(-\ell _2+\ell _3+1\right)\right) \Gamma
   \left(\frac{1}{2} \left(-\ell _3+\ell _4+1\right)\right) \Gamma
   \left(\frac{1}{2} \left(m-\ell _4+1\right)\right)}\,.
\eea
\end{widetext}
As a check, we have
\beq
P_{\vec{0}}=\frac{1}{\pi^\frac{3}{2}}\,,
\eeq
which corresponds to $1/\sqrt{{\rm Vol}(S_5)}$ with ${\rm Vol}(S_5)=\pi^3$.

The values
\beq
\mathcal{Y}_k=\sum_{ \vec{m}}m^k\left|Y_{\ell \vec{m}}\left(\frac{\pi}{2},\frac{\pi}{2},\frac{\pi}{2},\frac{\pi}{2},0\right)\right|^2\,,
\eeq
are listed in Table \ref{tab:1} below. In closed form such coefficients can be represented as
\bea
{\mathcal Y}_{2k}&=&\frac{4^{k-1} (\ell+2)}{\pi ^3} \left(\frac{(\ell+2)^2 B_{2
   k+1}\left(\frac{\ell}{2}+2\right)}{2 k+1}\right.\nonumber\\
   &-&\left.\frac{4 B_{2
   k+3}\left(\frac{\ell}{2}+2\right)}{2 k+3}\right)\,,
\eea
where $B_{n}(x)$ are the Bernoulli polynomials.

\begin{table*}  
\caption{\label{tab:1}  List of the values $\mathcal{Y}_{2k}$, for $k=0,\ldots 5$.
}
\begin{ruledtabular}
\begin{tabular}{ll}
$\mathcal{Y}_0$ & $\frac{(\ell+1)(\ell+2)^2(\ell+3)}{12\pi^3}$\\
$\mathcal{Y}_2$ & $\frac{\ell  (\ell +1) (\ell +2)^2
   (\ell +3) (\ell +4)}{60 \pi^3}$\\
$\mathcal{Y}_4$& $  \frac{\ell  (\ell +1) (\ell +2)^2
   (\ell +3) (\ell +4) \left(3
   \ell ^2+12 \ell -8\right)}{420
   \pi ^3}$ \\
$\mathcal{Y}_6 $ & $   \frac{\ell  (\ell +1) (\ell +2)^2
   (\ell +3) (\ell +4) \left(5
   \ell ^4+40 \ell ^3+40 \ell
   ^2-160 \ell +96\right)}{1260
   \pi ^3} $\\
$\mathcal{Y}_{8}$& $  \frac{\ell  (\ell +1) (\ell +2)^2
   (\ell +3) (\ell +4) \left(5
   \ell ^6+60 \ell ^5+160 \ell
   ^4-320 \ell ^3-768 \ell ^2+2048
   \ell -1152\right)}{1980 \pi ^3}$\\
 $\mathcal{Y}_{10}$& $     \frac{\ell  (\ell +1) (\ell +2)^2
   (\ell +3) (\ell +4)\Big(105
   \ell ^8+1680 \ell ^7+7280 \ell
   ^6-6720 \ell ^5-73920 \ell
   ^4+89600 \ell ^3+ 342400 \ell
   ^2-780800 \ell
   +421376\Big)}{60060 \pi ^3} $\\
  \end{tabular}
\end{ruledtabular}
\end{table*}

Finally, introducing the operator
\bea
\mathcal{T}_r&=& \sqrt{H}\mathcal{L}_r\nonumber\\
&=& \partial_r^2+\frac{5}{r}\partial_r+\Bigg[\omega^2H(r)-\frac{\ell(\ell+4)}{r^2}\Bigg]
\eea
we obtain
\bea\label{sourceLr}
\mathcal{T}_r(R(r))&=& 
\mathcal{S}\delta(r-r_c)\,,
\eea
with
\bea
\mathcal{S}&=& 
-\frac{8\pi^2q_{D3}}{\Gamma r_c^5}\delta(\omega-m\Omega)P_{\ell \ell_2 \ell_3 \ell_4 m}\nonumber\\
&\equiv & \hat {\mathcal S}\delta(\omega-m\Omega)P_{\ell \ell_2 \ell_3 \ell_4 m}\,.
\eea

The previous equation can be solved by using the Green function method
\beq
\mathcal{T}_r(G(r,r'))=\frac{1}{r'^5}\delta(r-r')\,,
\eeq
where
\beq
G(r,r')=\frac{1}{W}\Big[R_{\rm in}(r)R_{\rm up}(r')\Theta(r'-r)+ r\to r' \Big]\,,
\eeq
and where 
\beq
W=r^5(R_{\rm in}(r)R'_{\rm up}(r)-R'_{\rm in}(r)R_{\rm up}(r))
\eeq
is the constant wronskian.
For example, for the PN solution we find
\bea
W_{\rm PN}&=&-2 (\ell +2)+\eta^4\Bigg[-\frac{L^4 \omega ^4 \log
   \left(\frac{L \omega
   }{2}\right)}{(\ell+1)(\ell+3)}\nonumber\\
   &-&\frac{L^4 \omega ^4 (\ell
   +2) (\ell  (\ell +4)+1)}{4
   (\ell +1)^2 (\ell +3)^2}\Bigg]+O\left(\eta^8\right)\qquad
\eea

The solution of the non-homogeneous problem is
\bea
R(r)&=&\int dr' r'^5G(r,r')\mathcal{S}\delta(r'-r_c)\nonumber\\
&=&r_c^5G(r,r_c)\mathcal{S}\,.
\eea
In order to simplify the Fourier transform in \eqref{phidec}, let us introduce
\beq
R(r)=\mathfrak{R}(r)2\pi\delta(\omega-m\Omega)\,.
\eeq
Because of the limit $r\to \infty$ we can use only the up part of the Green function 
\bea\label{834}
\mathfrak{R}(r)&=& -\frac{4\pi q_{D3}}{\Gamma}P_{\ell \ell_2 \ell_3 \ell_4 m}\frac{R_{\rm up}(r)R_{\rm in }(r_c)}{W}\nonumber\\
&=& -\frac{4\pi q_{D3}}{\Gamma}P_{\ell \ell_2 \ell_3 \ell_4 m}\frac{R_{\rm in }(r_c)}{W} C_{\rm trans} \frac{e^{i\omega r}}{r^{5/2}}
\eea
where
\beq
 C_{\rm trans} =\frac{(i-1)}{\sqrt{\pi \omega}}\sum_{n=-\infty}^\infty e^{-\frac12 i \pi (\nu+n)} a_n
\eeq
For example,
\begin{widetext}
\bea
 C_{\rm trans}^{\ell=0} &=& \frac{(i-1)}{\sqrt{\pi \omega}} \left[ \frac{2}{9} \left(5+2 i \sqrt{5}\right)\right.\nonumber\\
 &+& \eta ^4 L^4 \omega ^4 \left(\frac{\left(460-383 i \sqrt{5}\right)
  }{25920}-\frac{5}{432} \left(\sqrt{5}+2 i\right) \pi  \right)\nonumber\\
&+& \left.   
 \eta ^8L^8 \omega^8 \left(\frac{\left(39856+14209 i \sqrt{5}\right) }{29859840}+\frac{5 \left(5+2 i \sqrt{5}\right) \pi ^2  }{82944}+\frac{\left(-26 \sqrt{5}+137 i\right) \pi  }{165888}\right)
 +O(\eta^{12})
\right]\,,\nonumber\\
 C_{\rm trans}^{\ell=1} &=& \frac{(i+1)}{\sqrt{\pi \omega}}
 \left[1+  \eta ^4 L^4 \omega^4\left(-\frac{5 }{256}+\frac{1}{192} i \pi \right) \right.\nonumber\\
&+&\left.  \eta ^8 L^8 \omega^8\left(\frac{63 }{819200}-\frac{\pi ^2}{73728}-\frac{179 i \pi  }{1105920}\right)
+O(\eta^{12})
\right]\,.
\eea

\end{widetext}

The energy flux \eqref{dedt} at infinity can be written as
\bea\label{fluxinf}
\frac{dE^{\infty}}{dt}=\lim_{r\to \infty}\frac{r^5}{4\pi}\sum_{\ell m}m\Omega {\mathcal F}_{\ell m}\,,
\eea
with
\bea
 {\mathcal F}_{\ell m}&=& \mathfrak{I}m\Bigg[\mathfrak{R}(r)\frac{d}{dr}\mathfrak{R}^*(r)\Bigg]_{\omega=m\Omega}\nonumber\\
 &\approx & \frac{16\pi^2 q_{D3}^2}{\Gamma^2}P_{\ell \ell_2 \ell_3 \ell_4 m}^2 \frac{|R_{\rm in}(r_c)|^2}{|W|^2} m\Omega \frac{|C_{\rm trans}|^2}{r^5}\,,\qquad
\eea
(here $\approx$ stands for the limiting behaviour at infinity) 
so that 
\bea
\frac{dE^\infty}{dt}= \frac{4\pi  q_{D3}^2 \Omega^2 }{\Gamma^2}\sum_{\ell m}m^2P_{\ell \ell_2 \ell_3 \ell_4 m}^2 \frac{|R_{\rm in}(r_c)|^2}{|W|^2}|C_{\rm trans}|^2\,, \qquad
\eea
Expanding in small $u=L/r_c$, we find
\begin{widetext}
\bea
\label{dE_dt_u}
\frac{dE^\infty}{dt}&=&\left(\frac{dE^\infty}{dt}\right)_N \left[ 
1+\frac{87 u^4}{8}+\frac{1559 u^8}{320}
+u^{12}
   \left(-\frac{\log (u)}{24}-\frac{\gamma
   }{48}-\frac{2383759}{6400}+\frac{\log (2)}{48}\right)\right.\nonumber\\
&+& u^{16} \left(-\frac{3811
   \log (u)}{960}-\frac{3811 \gamma
   }{1920}+\frac{446765119}{204800}-\frac{19 \log (2)}{128}\right) 
\nonumber\\
&+&\left. u^{20}
   \left(-\frac{8837 \log (u)}{320}-\frac{8837 \gamma
   }{640}-\frac{8466934987853}{1032192000}-\frac{1594323 \log
   (3)}{51200}+\frac{101529\log (2)}{3200}\right)
+O(u^{24}) \right]\,,
   \eea
with
\beq
\left(\frac{dE^\infty}{dt}\right)_N= \frac{ q_{D3}^2}{(4\pi)L^5 48 }u^{19}\,.
\eeq
Rewriting Eq. \eqref{dE_dt_u} as a function of the gauge invariant variable $y_\Omega$ defined in Eq. \eqref{y_Omega}  gives the following gauge-invariant expression
\bea
\frac{dE^\infty}{dt}\bigg|_{y_\Omega}&=&\left(\frac{dE^\infty}{dt}\right)_{N, y_\Omega} \left[ 
1+\frac{337 y}{24} +\frac{50017 y^2}{960}
+\left(-\frac{1}{96} \log \left(e^{2 \gamma } y\right)-\frac{133086869}{518400}+\frac{\log (2)}{48}\right)y^3 \right.\nonumber\\
&+&  \left(-\frac{12053
   \log \left(e^{2 \gamma }
   y\right)}{11520}+\frac{22181489453}{49766400}-\frac{47 \log
   (2)}{1152}\right)y^4\nonumber\\
&+& \left. \left(-\frac{295861 \log \left(e^{2 \gamma }
   y\right)}{23040}-\frac{1621917613999}{3096576000}-\frac{1594323 \log
   (3)}{51200}+\frac{1794697 \log (2)}{57600}\right)y^5    
+O(y^{6}) \right]\,,
\eea
where we have conveniently collected the Euler gamma terms inside the logarithms and
\beq
\left(\frac{dE^\infty}{dt}\right)_{N, y_\Omega}=\frac{ q_{D3}^2}{(4\pi) L^5 48 }y^{19/4}\,.
\eeq
\end{widetext}
From \eqref{fluxinf} and considering the behaviour near the horizon $r=0$ rather than $r\to \infty$, the ingoing flux at the horizon is
\bea
\frac{dE^H}{dt}=\frac{4\pi q_{\rm D3}^2L^2\Omega^2}{\Gamma^2}\sum_{\ell \vec{m}}\frac{|B_{\rm trans}|^2|R_{\rm up}(r_c)|^2}{|W|^2}P^2_{\ell \vec{m}} m^2\,,\qquad
\eea
where
\bea
B_{\rm trans}=\frac{(-1)^\frac{3}{4}}{L\sqrt{2\pi\omega}}\sum_{n=-\infty}^\infty e^{-\frac{i\pi}{2}(n+\nu)}(e^{2\pi i(n+\nu)}-1)a_n\,.\qquad
\eea
For example
\bea
B_{\rm trans}^{\ell=0}&=&\frac{5}{216}(i-1)(2i+\sqrt{5})L^3\sqrt{\pi}\omega^\frac{7}{2}\eta^4\nonumber\\
&-&\frac{1}{82944}(i+1)\Big(20 \left(2
   \sqrt{5}-5 i\right) \pi-137\nonumber\\
   &-&26 i \sqrt{5}\Big)L^7\sqrt{\pi}\omega^\frac{15}{2}\eta^8+O\left(\eta^{12}\right)\nonumber\\
B_{\rm trans}^{\ell=1}&=&\frac{1}{96}(1-i)L^3\sqrt{\pi}\omega^\frac{7}{2}\eta^4\nonumber\\
&-&\frac{1}{552960}(1+i)(30\pi-179i)L^7\sqrt{\pi}\omega^\frac{15}{2}\eta^8\nonumber\\
&+&O\left(\eta^{12}\right)\,.\qquad
\eea
The final result including contributions from the $\ell=0,1,2,3$ multipoles in the dimensionless variable u is
\beq
\frac{dE^H}{dt}=\left(\frac{dE^H}{dt}\right)_N\left(1-5u^4+30u^8+O(u^{12})\right)
\eeq
where the Newtonian contribution is
\beq
\left(\frac{dE^H}{dt}\right)_N=\frac{q_{\rm D3}^2}{L^5\pi}\frac{u^{31}}{192}\,.
\eeq
In terms of the variable $y$ the energy flux at horizon reads
\bea
\frac{dE^H}{dt}\bigg|_{y_\Omega}&=&\left(\frac{dE^H}{dt}\right)_{N, y_\Omega} \Big(1+\frac{y}{6}+\frac{361}{24}y^2+O\left(y^3\right)\Big)\,,\nonumber\\
\left(\frac{dE^H}{dt}\right)_{N, y_\Omega}&=&\frac{q_{\rm D3}^2}{L^5\pi}\frac{y^\frac{31}{4}}{192}\,,
\eea
and turns out to be strongly PN-suppressed with respect to outgoing flux at infinity.

We conclude this section by recalling that, although this computation may seem straightforward with modern computational tools, this is true only in four dimensions. Extending the analysis beyond the relative \lq\lq safety\rq\rq of four dimensions remains a challenging task, with potentially important implications for future applications. For example, this issue is expected to be particularly relevant in the context of dimensional regularization for four-dimensional spacetimes.

\section{Conclusions}

D3 brane solutions provide a relatively simple setting in which to carry out analytical calculations of scalar self-force effects in higher-dimensional spacetimes.
The case considered here is based on a ten-dimensional, spherically symmetric background (with a number of geometrical information collected, mainly as a reminder). In particular, the angular variables parametrize the five-sphere  $S^5$, and the corresponding angular dependence is characterized by a set of five quantum numbers (which must be summed over to reconstruct the associated functions).

We have provided here explicit calculations of perturbations due to  a scalar field sourced by a scalar charge moving on a (properly defined) circular orbit.

In particular, we have computed the energy flux at infinity in PN-expanded form and investigated the flux entering the horizon, finding it to be strongly suppressed in the near-horizon limit. They constitute key gauge-invariant observables and represent one of the main results of the present study.

To get the final result mentioned above, we have extended the MST technology (poorly studied out of the four dimensional black hole case) to this context, i.e., another nontrivial outcome of the present study, necessary to construct the solutions of the radial equation with proper boundary conditions. 
The latter radial equation turns out to be a source-modification of the well-known Mathieu equation. Unfortunately, this circumstance does not involve particular simplifications.

Finally, further studies in this context should involve PN-resummations or possible ways of getting strong field information, open issues to be addressed in the next years.

\section*{Acknowledgments}

D.B. and G.d.R. acknowledge  membership to the Italian Gruppo Nazionale per
la Fisica Matematica (GNFM) of the Istituto Nazionale
di Alta Matematica (INDAM).
V.F. is grateful to the Istituto
per le Applicazioni del Calcolo \lq\lq M. Picone," CNR, Rome (IT) for hospitality during the development of
the present project.
We thank 
Ilia Gaiur and Frédéric Menous 
for many useful discussions.

\appendix

\section{Gegenbauer polynomials}\label{appa}

The Gegenbauer polynomials, used here to represent the spherical harmonics on $S_5$, satisfy the following differential equation
\beq
(1{-}x^2)\frac{d^2 G_n^\lambda(x)}{dx^2}{-}(2\lambda{+}1)x\frac{d G_n^\lambda(x)}{dx}{+}n(n{+}2\lambda) G_n^\lambda(x){=}0\,.
\eeq
Furthermore the following equation holds
\beq
\frac{d G_n^\lambda(x)}{dx}=2\lambda G_{n-1}^{\lambda+1}(x)\,.
\eeq
The orthogonality of the Gegenbauer polynomials is
\beq
\int_{-1}^1 dx(1-x^2)^{\alpha-\frac{1}{2}} G_n^\lambda(x)G_m^\lambda(x)=g_{n,\alpha} \delta_{nm}\,,
\eeq
where
\beq
g_{n,\alpha}=\frac{\pi 2^{1-2\alpha}\Gamma(n+2\alpha)}{n!(n+\alpha)\Gamma(\alpha)^2}\,.
\eeq

\section{MST-like solution for $\ell=1$}\label{appb}
We display below as an example the first few terms of a PN-expanded form of the \lq\lq in" and \lq\lq up" solutions of the radial equation for $\ell=1$, computed by using a straightforward generalization of the MST technology (original with this work). 

\begin{widetext}
    \bea
 R_{\rm in}^{\ell=1}&{=}&{-}\frac{r \omega }{6 L^2}{+}\eta^2\Big[\frac{r^3 \omega ^3}{96
   L^2}{-}\frac{L^2 \omega ^3}{48
   r}\Big]{+}\eta^4\Big[{-}\frac{L^6 \omega ^5}{384
   r^3}{-}\frac{r^5 \omega
   ^5}{3840 L^2}{-}\frac{({-}269{+}60
   \gamma {-}60 i \pi ) L^2 r
   \omega
   ^5}{34560}{-}\frac{1}{576} L^2
   r \omega ^5 \log
   \left(\frac{L^2 \omega }{2
   r}\right)\Big] \nonumber\\
   &+&\eta^6\Big[\frac{(-11+12 \gamma -6 i \pi )
   L^{10} \omega ^7}{27648
   r^5}-\frac{(-97+30 \gamma
   -30 i \pi ) L^6 \omega
   ^7}{138240 r}+\frac{r^7
   \omega ^7}{276480
   L^2}+\frac{(-89+24 \gamma
   -24 i \pi ) L^2 r^3 \omega
   ^7}{221184}\nonumber\\
   &+&\left(\frac{L^{1
   0} \omega ^7}{2304
   r^5}-\frac{L^6 \omega
   ^7}{4608 r}+\frac{L^2 r^3
   \omega ^7}{9216}\right) \log
   \left(\frac{L^2 \omega }{2
   r}\right)\Big]+O\left(\eta^8\right)\,,\nonumber\\
    R_{\rm up}^{\ell=1}&=&-\frac{16 i}{\pi  r^5 \omega ^3}+\eta^2\Bigg[\frac{i \left(L^4-2
   r^4\right)}{\pi  r^7 \omega }\Bigg]+\eta^4\Bigg[-\frac{i L^4 \omega  \log
   \left(\frac{r \omega
   }{2}\right)}{6 \pi 
   r^5}-\frac{i \omega  \left(6
   L^8+5 (8 \gamma -27) L^4
   r^4+60 r^8\right)}{240 \pi 
   r^9}\Bigg]\nonumber\\
   &+&\eta^6\Bigg[\frac{i \omega ^3 \left(L^8-2 L^4
   r^4+4 r^8\right) \log
   \left(\frac{r \omega
   }{2}\right)}{96 \pi 
   r^7}\nonumber\\
   &+&\frac{i \omega ^3 \left(2
   L^{12}+12 (5 \gamma -13) L^8
   r^4+15 (17-8 \gamma ) L^4
   r^8+20 (-11+12 \gamma -6 i \pi
   ) r^{12}\right)}{5760 \pi 
   r^{11}}\Bigg]+O\left(\eta^8\right)\,.
\eea
\section{Energy flux in the decoupling limit}
Let's rewrite Eq. \eqref{820} in term of the variable $y=\frac{4r^2}{L^4\omega^2}$ and rescale the wavefunction as $R(y)=\psi(y) y^{-\frac{3}{2}}$, so that
\beq
\psi''(y)+Q_W(y)\psi(y)=-\frac{\pi L^4 q_{\rm D3}\sqrt{r_c^4-L^4}\sqrt{y}\omega^2}{2r_c^6(r_c^4+L^4)^\frac{1}{4}}\delta(\omega-m \Omega)\delta(2r_c-L^2\omega \sqrt{y})Y^{*}_{\ell\vec m}\left(\frac{\pi}{2},\frac{\pi}{2},\frac{\pi}{2},\frac{\pi}{2},\Omega t\right)\,,
\eeq
\end{widetext}
where
\beq
Q_W(y)=\frac{1}{y^3}-\frac{(1+\ell)(3+\ell)}{4y^2}+\frac{L^4\omega^4}{16y}
\eeq
is the effective potential. The previous function can be expanded at $y\to\infty$ as
\beq
Q_W(y)\underset{y\to\infty}{\sim}\frac{L^4\omega^4}{16y}-\frac{(\ell+1)(\ell+3)}{4y^2}\,.
\ee
In presence of such potential, the homogeneous equation can be solved exactly in terms of Bessel functions. 
\bea
\psi(y)&=&\sqrt{y}\Big[c_1 I_{\ell +2}\left(\frac{1}{2}
   i L^2  \omega
   ^2\sqrt{y}\right)\nonumber\\
   &+&c_2 i^{2 \ell }
   K_{\ell +2}\left(\frac{1}{2} i
   L^2 \omega ^2\sqrt{y} \right)\Big]\,.
\eea
In order to select the solution with outgoing behavior at infinity, we must choose the constants as follows
\beq
c_2=-\frac{ic_1}{\pi}\,,
\eeq
so that the \lq\lq up" solution in the decoupling limit is
\bea
\psi_{\rm up}(y)&=&\sqrt{y}\Big[\pi  I_{\ell +2}\left(\frac{1}{2}
   i L^2 \omega
   ^2\sqrt{y} \right)\nonumber\\
   &-& i^{2 \ell +1}
   K_{\ell +2}\left(\frac{1}{2} i
   L^2 \omega ^2\sqrt{y} \right)\Big]\,.
\eea
Around $y=0$ the regular behavior is selected requiring $c_2=0$ and so
\beq
\psi_{\rm in }(y)=\sqrt{y} I_{\ell +2}\left(\frac{1}{2} i
   L^2 \omega ^2\sqrt{y} \right)\,.
\eeq
Given $\psi_{\rm in }(y)$ and $\psi_{\rm up }(y)$ the constant Wronskian is
\beq
W=\frac{i^{2\ell+1}}{2}\,.
\eeq
The particular solution  \eqref{834} is
\bea
\mathfrak{R}(r)&=&e^{i\omega r}\sqrt{r}\frac{(1+i)i^{-2\ell}\sqrt{\pi}q_{\rm D3}\sqrt{r_c^4-L^4}}{r_c^3(r_c^4+L^4)^\frac{1}{4}\sqrt{\omega}}\nonumber\\
&\times&I_{\ell +2}\left(i \omega 
   r_c\right) Y^{*}_{\ell\vec m}\left(\frac{\pi}{2},\frac{\pi}{2},\frac{\pi}{2},\frac{\pi}{2},\Omega t\right)\,.
\eea
So the (decoupling limit) energy flux is in small $L/r_c$ expansion
\bea
\frac{dE}{dt}&=&\frac{41q_{\rm D3}^2}{10080\pi^3r_c^5}z^{10}\Big[1-\frac{1095489 z^4}{461824}\nonumber\\
&+&\frac{170640651 z^8}{24014848}-\frac{20132211461
   z^{12}}{8645345280}\nonumber\\
   &-&\frac{28834884487
   z^{16}}{14251720704}\Big]+O(z^{30})\,,
\eea
where $z=L/r_c$.

\bibliographystyle{apsrev}
\bibliography{references}
\end{document}